\documentclass[prx,aps,twocolumn,groupedaddress,floatfix]{revtex4-2}

\usepackage[pdftex]{graphicx}
\usepackage{amsbsy,amssymb,amsmath,bm,mathtools}

\usepackage{xspace}

\usepackage{color}
 
\usepackage[mathscr]{euscript}

\usepackage{url}
\usepackage[section]{placeins}
\usepackage[colorlinks=true,linkcolor=blue,citecolor=blue]{hyperref}

\begin{document}

\raggedbottom

\title{Exact Equivariance from Ordinary Neural Networks for Lattice Many-Body Dynamics}

\author {Ho Jang}
\affiliation{Department of Physics, University of Virginia, Charlottesville, Virginia, 22904, USA}

\author {Sankha Subhra Bakshi}
\affiliation{Department of Physics, University of Virginia, Charlottesville, Virginia, 22904, USA}

\author {Gia-Wei Chern}
\affiliation{Department of Physics, University of Virginia, Charlottesville, Virginia, 22904, USA}

\date{\today}

\begin{abstract}
Large-scale simulations of correlated electron systems require repeated evaluations of the electronic forces and transition energies driving collective dynamics. Machine-learning surrogates alleviate this bottleneck, but incorporating symmetry often involves carefully designed descriptors or specialized network architectures. We show that ordinary multilayer perceptrons equipped with finite-group averaging provide exactly equivariant surrogates directly from microscopic configurations. The construction separates symmetry enforcement from the internal network architecture and applies to both discrete and continuous lattice degrees of freedom. In the Falicov--Kimball model, it predicts directional hopping free-energy differences; in the Holstein model, its invariant limit generates conservative lattice forces. Benchmarks against exact diagonalization establish microscopic accuracy and agreement of dynamical correlations, while large-scale simulations recover multiscale charge ordering and charge-density-wave coarsening. These results demonstrate an accessible, reusable route to symmetry-preserving many-body dynamics for systems with finite lattice point groups.
\end{abstract}

\maketitle

\section{Introduction}
\label{sec:introduction}

Spatially inhomogeneous states arise throughout correlated electron systems, from charge stripes, electronic liquid-crystal and intertwined orders, pair-density waves, and coexisting metallic and insulating phases to noncollinear spin order and magnetic skyrmions~\cite{dagotto05,kivelson98,fradkin15,hamidian16,qazilbash07,mcleod17,Tokura2017,nagaosa13,Kurumaji2019}. Their formation and evolution reflect the interplay between microscopic electronic interactions and collective order-parameter fields over a wide range of length and time scales. Understanding processes such as domain growth, competition between nearly degenerate orders, and the relaxation or coarsening of complex textures therefore requires real-space simulations that can access systems much larger and time scales much longer than those typically available to direct electronic calculations~\cite{hohenberg77,bray94,furukawa85}. A central difficulty is that the effective forces, local energy changes, and transition rates governing these collective degrees of freedom are themselves determined by the underlying fermions. Repeated solution of the electronic problem can consequently become the dominant computational bottleneck, even when the microscopic Hamiltonian is tractable for any fixed configuration.

Machine learning (ML) provides a natural route around this bottleneck by replacing repeated electronic calculations with efficient surrogate models. Machine-learned interatomic potentials exemplify this strategy, reproducing potential-energy surfaces and atomic forces from electronic-structure training data at greatly reduced computational cost and thereby extending first-principles accuracy to much larger-scale simulations~\cite{behler07,bartok10,li15,smith17,chmiela17,zhang18,deringer19,noe20,suwa19}. Related ideas have more recently been applied to correlated lattice systems, enabling large-scale simulations of phase separation and charge ordering, orbital ordering in Jahn--Teller systems, and nonequilibrium spin dynamics in itinerant magnets~\cite{zhang21,zhang22b,cheng23a,Ghosh24,cheng23b,zhang23,Fan24,tyberg25,Chern2026b}. In these settings the learned object need not be a scalar energy or a conventional force. More generally, one seeks a map from a microscopic configuration $X$ to a physical quantity $Y$---for example a local energy, a generalized force, a transition-energy vector, or another dynamical quantity---while preserving the symmetry relations between the corresponding input and output representations.

An essential requirement is that such predictions respect the symmetries of the microscopic problem. Scalar energies must be invariant, whereas outputs carrying spatial, directional, or internal indices must transform equivariantly. For invariant targets, a successful strategy is to encode the local environment using symmetry-invariant descriptors and supply these features to a conventional learning model~\cite{behler11,bartok13,ghiringhelli15,himanen20,huo22,drautz19,zhang22}. This approach cleanly separates symmetry handling from the regression problem and has proved highly effective in both atomistic and lattice applications. Its practical difficulty is that an informative descriptor must retain enough information to distinguish physically inequivalent environments while discarding only symmetry-related redundancy. Constructing such descriptors can require substantial system-specific choices involving basis functions, correlation order, reference environments, or feature truncation. The problem becomes still more involved for directional outputs, because the representation must retain enough orientation information to reconstruct the required transformation law rather than eliminate it altogether.

Equivariant neural networks address this issue by incorporating prescribed transformation laws directly into the architecture, often together with graph-based message passing~\cite{cohen2016,cohen2018,thomas2018,weiler2018,Geiger22,batzner2022,batatia2022,musaelian2023}. By organizing intermediate features into symmetry representations and combining them through operations such as group convolutions and tensor products, these methods can learn highly expressive symmetry-adapted representations and have achieved remarkable success, particularly for systems with continuous spatial symmetries. For many lattice problems, however, the situation is considerably simpler. The geometry is fixed, the relevant point group is finite and usually small, and each symmetry operation acts as a known permutation or linear transformation of the microscopic variables. This raises a more elementary possibility: rather than redesigning the hidden layers of the network, can one impose the exact transformation law directly on the input-output map while leaving the internal learning model completely unrestricted?

Here we introduce an equivariant projection network (EPN) framework that realizes precisely this idea for lattice many-body dynamics. An EPN evaluates an ordinary multilayer perceptron (MLP) on all symmetry-related versions of a local microscopic configuration, transforms the corresponding predictions back to a common frame, and averages them. The resulting map is exactly equivariant by construction, independently of the parameters or architecture of the underlying MLP. Thus symmetry is imposed at the level of the physical input-output relation rather than through symmetry-engineered descriptors or equivariant hidden features. The underlying group-averaging construction has precedents in frame averaging, equivariant fine-tuning, Reynolds networks, and related applications including lattice-Boltzmann models~\cite{puny2022,basu2023a,sannai2024,dittmer2022,corbetta2023,duval2023}. Our emphasis here is on its particularly direct use for fixed-lattice many-body problems, where the relevant symmetry action is known exactly and the full finite-group average is inexpensive.

We develop this construction into a general surrogate framework for fermion-mediated dynamics, where repeated electronic calculations otherwise limit accessible system sizes and evolution times. The learned maps may represent observables, generalized forces, local transition energies, or other quantities entering a dynamical update, with the appropriate transformation law imposed by the same finite-group projection. We first consider Falicov--Kimball dynamics, where a local hop is characterized by four directional free-energy differences and therefore provides a nontrivial equivariant output. We then consider the Holstein model, where the invariant special case of the construction is combined with the Behler--Parrinello (BP) decomposition of the total energy into local contributions~\cite{behler07,behler11}. Differentiation of the resulting invariant energy yields a symmetry-preserving and conservative force field. Together, these examples show that both invariant and directional lattice surrogates can be constructed from ordinary MLPs without invariant descriptors, canonical local frames, or symmetry-constrained hidden representations.

The Falicov--Kimball and Holstein models are useful here because closely related dynamics have already been studied with descriptor-based ML surrogates~\cite{zhang22b,zhang22,cheng23a}. They provide complementary benchmarks for assessing whether direct symmetry projection can reproduce microscopic electronic calculations and established collective dynamics with substantially simpler input processing. Section~\ref{sec:EPN} presents the general construction, followed by the directional FK application in Sec.~\ref{sec:FK} and the invariant-energy Holstein application in Sec.~\ref{sec:Holstein}. We discuss the scope and practical advantages of the framework in Sec.~\ref{sec:discussion}.

\section{Equivariant projection networks}
\label{sec:EPN}

\begin{figure*}
    \centering
    \includegraphics[width=0.99\linewidth]{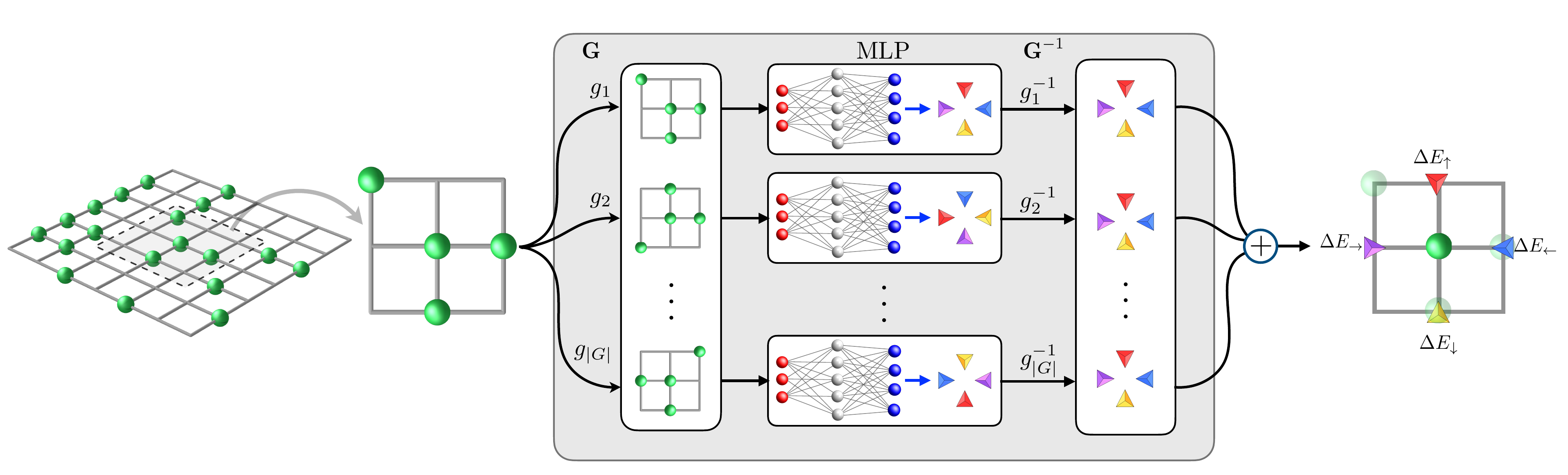}
\caption{Equivariant projection network (EPN) for Falicov--Kimball dynamics. Each $D_4$ operation transforms the local $f$-electron configuration before it enters a shared MLP. The four predicted hopping free-energy differences are transformed back and averaged, yielding an exactly equivariant output.}
    \label{fig:FK-schematic}
\end{figure*}

To formulate the EPN, let $X_i$ denote the configuration in a finite neighborhood of site $i$ and $Y_i$ a local target obtained from a microscopic calculation. Motivated by electronic nearsightedness~\cite{kohn96,prodan05}, we approximate this calculation by $X_i\mapsto Y_i$; the accuracy of the finite cutoff must be checked for the regime of interest. The neighborhood is chosen to be closed under a finite symmetry group $G$, with input and output representations $D_X(g)$ and $D_Y(g)$. Equivariance requires
\begin{equation}
F\!\left[D_X(g)X\right]=D_Y(g)F(X).
\label{eq:equivariance}
\end{equation}
Starting from an unrestricted network $f$, we evaluate every symmetry-related input, transform its output back to the original frame, and average:
\begin{equation}
F(X)=\frac{1}{|G|}\sum_{g\in G}D_Y(g^{-1})\,f\!\left[D_X(g)X\right].
\label{eq:EPN}
\end{equation}
The inverse output transformation is essential for a directional target. Each evaluation $f[D_X(g)X]$ predicts components in the orientation of the transformed input, and $D_Y(g^{-1})$ returns those components to the original frame. The average therefore combines predictions for the same physical directions. Simply averaging the untransformed outputs would instead produce an invariant map and generally erase the directional dependence one seeks to retain. Unlike data augmentation, which encourages symmetry through examples, the projection imposes the transformation law for every choice of network parameters, including before training.

An input symmetry merely relabels the terms in this average. Explicitly, for $h\in G$,
\begin{align}
F\!\left[D_X(h)X\right]
&=\frac{1}{|G|}\sum_{g\in G}D_Y(g^{-1})
f\!\left[D_X(gh)X\right] \nonumber\\
&=\frac{1}{|G|}\sum_{k\in G}D_Y(h)D_Y(k^{-1})
f\!\left[D_X(k)X\right] \nonumber\\
&=D_Y(h)F(X),
\label{eq:EPN-equivariance}
\end{align}
where $k=gh$. Thus the transformation law holds independently of the internal architecture and the trained network parameters. The average is a projection: it leaves every equivariant map unchanged. Consequently, universal MLP approximation of continuous maps on compact, $G$-invariant input domains carries over to the equivariant maps on those domains~\cite{cybenko89,hornik89,barron93,puny2022,sannai2024}. To see why expressivity is preserved, consider an MLP approximating a desired equivariant map. Applying the projection leaves that target unchanged and averages the transformed approximation errors. The symmetry constraint therefore does not exclude any continuous equivariant target from the approximation class.

Implementation requires only the input and output group actions and $|G|$ evaluations of a shared MLP; the group-related inputs can be batched. For the square-lattice point group $D_4$, this means eight evaluations per local prediction, a fixed overhead independent of the simulated lattice size. A scalar output has $D_Y(g)=1$ and requires only invariant averaging. In the examples below the backbone $f$ is an ordinary fully connected MLP with SiLU activations and a linear output layer, trained with AdamW. No hidden layer is symmetry constrained: the loss is evaluated on the projected prediction, and gradients simply propagate through the group average to the shared MLP parameters. Architecture, normalization, and optimization details are collected in the Appendices.

\section{Falicov--Kimball dynamics}
\label{sec:FK}

We first consider the spinless Falicov--Kimball (FK) model, a paradigmatic setting for charge ordering, segregation, and correlation-driven metal--insulator behavior~\cite{falicov69,kennedy86,freericks03,freericks00,freericks02,lemanski02}:
\begin{equation}
\mathcal{H}=-t\sum_{\langle ij\rangle}\left(c_i^\dagger c_j+\mathrm{H.c.}\right)
+U\sum_i n_i^c n_i^f ,
\label{eq:FK}
\end{equation}
Here $c_i^\dagger$ creates an itinerant electron, $n_i^c=c_i^\dagger c_i$, and $n_i^f=0,1$ denotes a localized-electron occupation on the square lattice. For a fixed configuration $\{n_i^f\}$, the $c$ electrons move in the potential landscape generated by the localized particles.

To describe the slow configurational dynamics, we allow thermally activated nearest-neighbor $f$-electron hops and assume an adiabatic separation of time scales: the itinerant $c$ electrons equilibrate rapidly for each instantaneous $f$-electron configuration~\cite{zhang22b}. For a proposed hop of an $f$ electron from an occupied site $i$ in direction $\alpha$, the relevant electronic quantity is the free-energy difference $\Delta E_{i,\alpha}$ between the configurations before and after the hop, evaluated at fixed itinerant-electron number. Hops to occupied destinations are excluded, while an allowed move is accepted with probability
\begin{equation}
P_{i,\alpha}
=
\min\left[1,\exp\left(-\frac{\Delta E_{i,\alpha}}{T}\right)\right].
\label{eq:FK-metropolis}
\end{equation}
Direct evaluation of $\Delta E_{i,\alpha}$ requires solving the fermionic problem for both configurations. Because such electronic calculations must be repeated for every attempted local update, they rapidly become the dominant computational cost of a large-scale kinetic Monte Carlo (kMC) simulation.

This problem was previously addressed using descriptor-based ML surrogates~\cite{zhang22b,zhang22}. There the local occupation pattern is first converted into symmetry-invariant features, while additional reference information is introduced to recover the directional character of a hop. In practice, the construction therefore requires more than choosing a local cutoff: the implementation may involve an invariant basis, relative-orientation information, reference configurations or canonical frames, and separate treatment of symmetry-inequivalent local patterns and their residual point groups. These descriptors can accurately reproduce the electronic transition energies, but the machinery is considerably more elaborate than the underlying symmetry statement: a lattice point-group operation simply permutes the sites of the local environment and, at the same time, permutes the possible hopping directions.

Figure~\ref{fig:FK-schematic} illustrates how the EPN implements this transformation law directly for the four directional hopping free-energy differences. For a selected occupied site $i$, let $\mathcal N$ denote the local neighborhood centered at $i$. The microscopic input is simply the ordered occupation vector
\begin{equation}
X_i=
\left(n^f_{i+\bm\delta}\right)_{\bm\delta\in\mathcal N},
\label{eq:FK-input}
\end{equation}
while the output has a fixed component for each of the four nearest-neighbor hopping directions,
\begin{equation}
Y_i=
\left(\Delta E_{i,\alpha}\right)_{\alpha\in\mathcal D},
\qquad
\mathcal D=\{+x,-x,+y,-y\}.
\label{eq:FK-output}
\end{equation}
In the calculations below, $\mathcal N$ is a disk of radius $R_c=10$, containing 317 sites including the center. A point-group operation $g\in D_4$ acts on $X_i$ through a fixed permutation of the occupation variables and on $Y_i$ through the corresponding permutation of the hopping directions. For example, a counterclockwise quarter-turn maps the $+x$ hopping direction to $+y$ while rotating the occupation pattern around the central site. The inverse output permutation returns the corresponding prediction to the $+x$ component before averaging. Equation~(\ref{eq:EPN}) thus combines eight predictions in a common directional frame. The output always has four components; a separate occupancy mask excludes hops into occupied sites from the training loss and the dynamical updates. The construction requires neither invariant descriptors nor a canonical local orientation.

\begin{figure}[t]
    \centering
    \includegraphics[width=0.99\columnwidth]{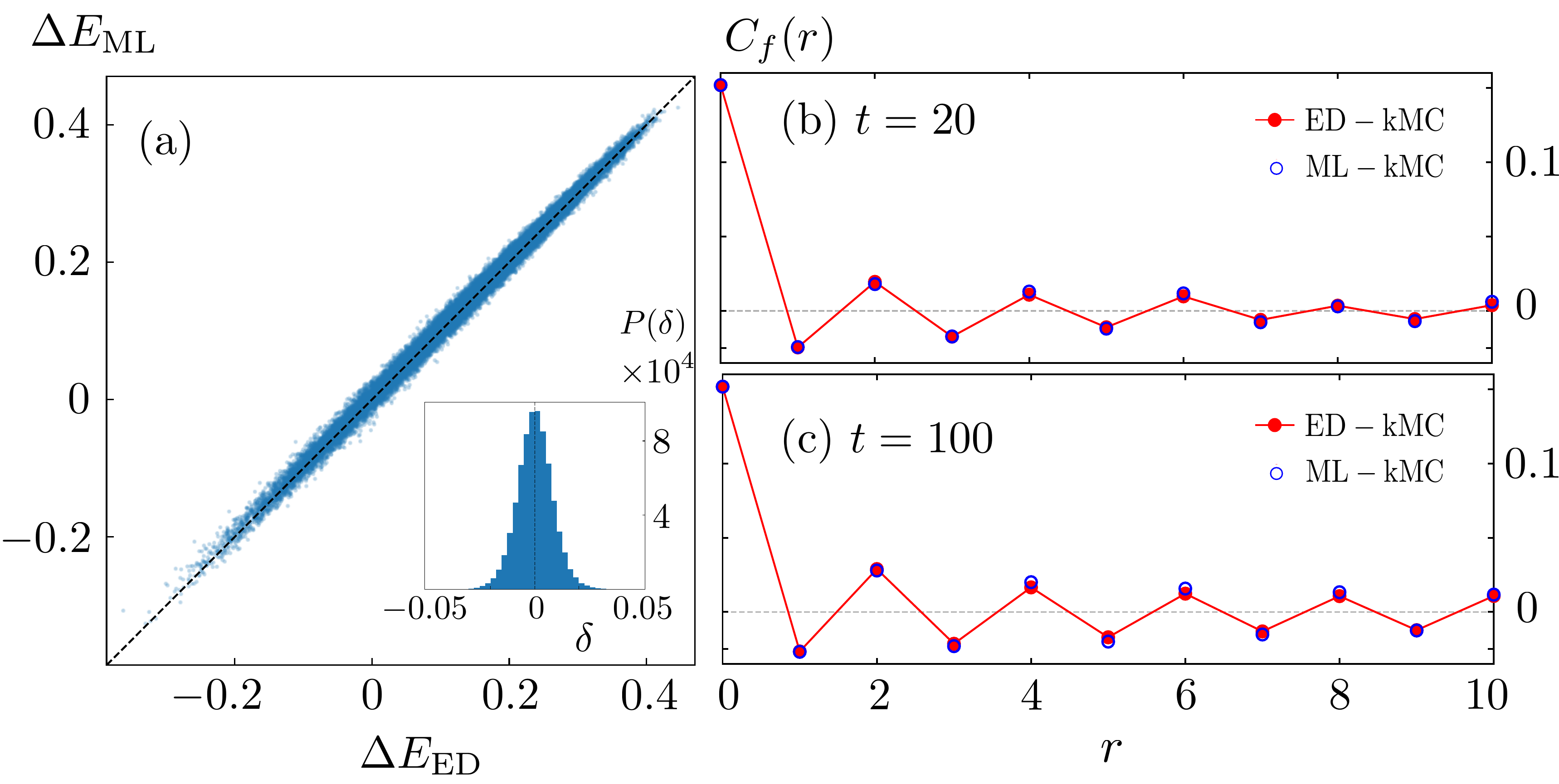}
\caption{Microscopic and dynamical benchmarks of the EPN for the Falicov--Kimball model. (a) Predicted hopping energy changes $\Delta E_{\rm ML}$ versus exact-diagonalization values $\Delta E_{\rm ED}$. The dashed line denotes perfect agreement; the inset shows the distribution of prediction errors $\delta=\Delta E_{\rm ML}-\Delta E_{\rm ED}$. (b,c) $f$-electron density correlation function $C_f(r)=\langle n_i^f n_j^f\rangle-\rho_f^2$, with $r=|\mathbf r_i-\mathbf r_j|$, obtained from ED-kMC and EPN-kMC simulations after a thermal quench at (b) $t=20$ and (c) $t=100$ sweeps.}
\label{fig:FK-benchmark}
\end{figure}

\begin{figure}[t]
    \centering
    \includegraphics[width=0.99\columnwidth]{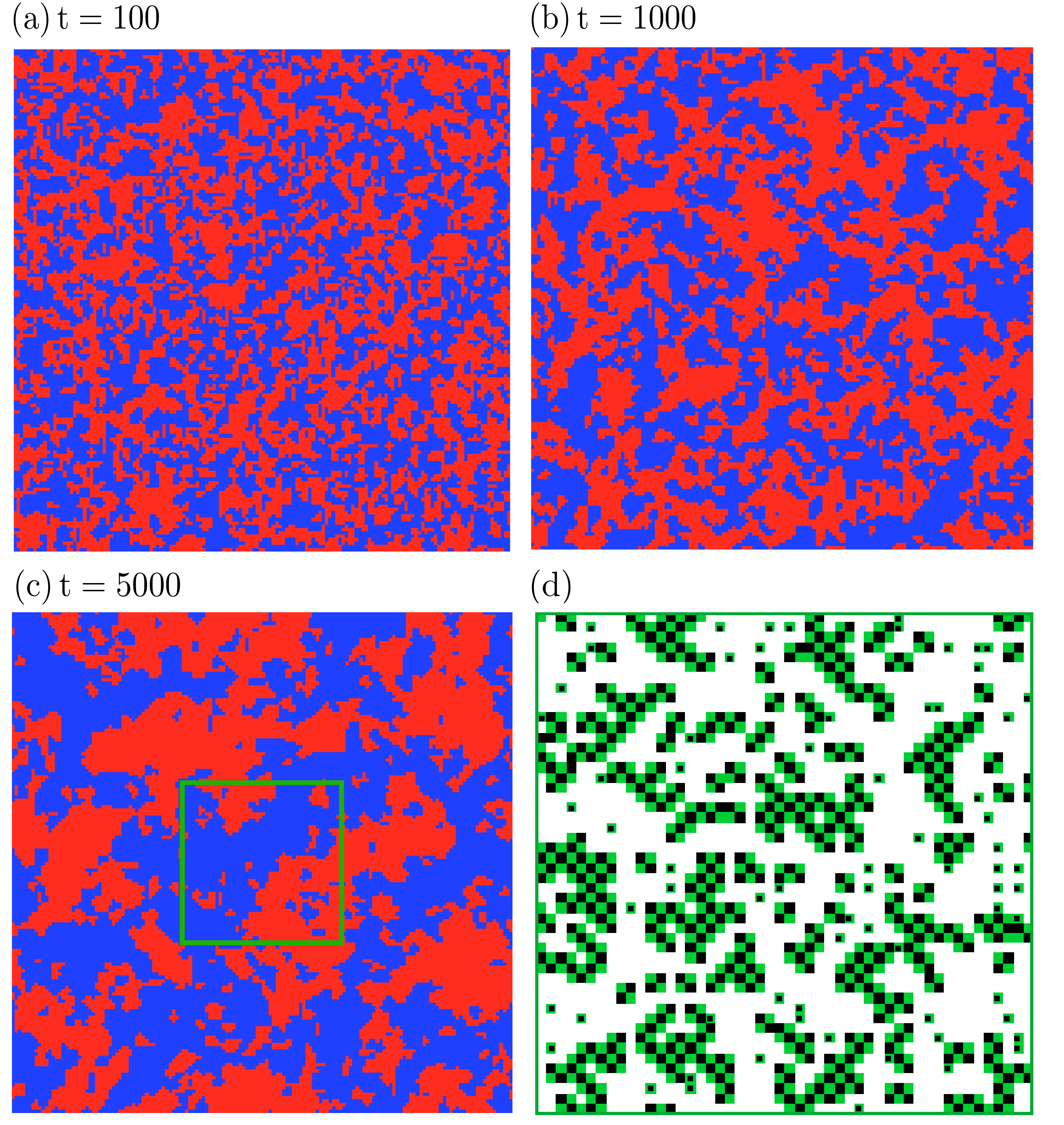}
\caption{Large-scale FK phase ordering with EPN-kMC. (a--c) Effective Ising configurations at $t=100$, $1000$, and $5000$ sweeps show coarsening of superclusters. Red and blue distinguish opposite sublattice polarities. (d) Microscopic $f$-electron configuration in the green square in (c), resolving the disconnected checkerboard clusters underlying the larger domains.}
    \label{fig:FK-large}
\end{figure}

For this benchmark we use a conventional MLP with hidden widths $512$, $512$, $256$, and $128$, followed by four linear outputs. The model is trained on 77,520 ED local environments, with 19,380 additional environments reserved for validation, using a masked normalized mean-square loss that includes only physically allowed hops. AdamW optimization is performed for 500 epochs with learning rate $10^{-3}$, weight decay $10^{-6}$, and batches of 1024 local environments; the complete definition is given in Appendix~\ref{app:FK-details}. These choices are not symmetry specific---the only symmetry operation is the EPN projection itself.

Figure~\ref{fig:FK-benchmark}(a) benchmarks the EPN directly against exact diagonalization using held-out local configurations. The predicted hopping free-energy differences $\Delta E_{\rm ML}$ closely follow the ED values $\Delta E_{\rm ED}$ over the full range shown, and the error distribution in the inset is narrow and centered near zero. This agreement shows that the microscopic occupation pattern within the cutoff contains sufficient information to accurately predict the energetics of a local hop. In particular, no symmetry-adapted features need to be constructed beforehand: the EPN operates directly on the ordered binary configuration, with the $D_4$ symmetry acting simply as permutations of the microscopic input and of the four directional outputs.

\begin{figure*}[t]
    \centering
    \includegraphics[width=1.99\columnwidth]{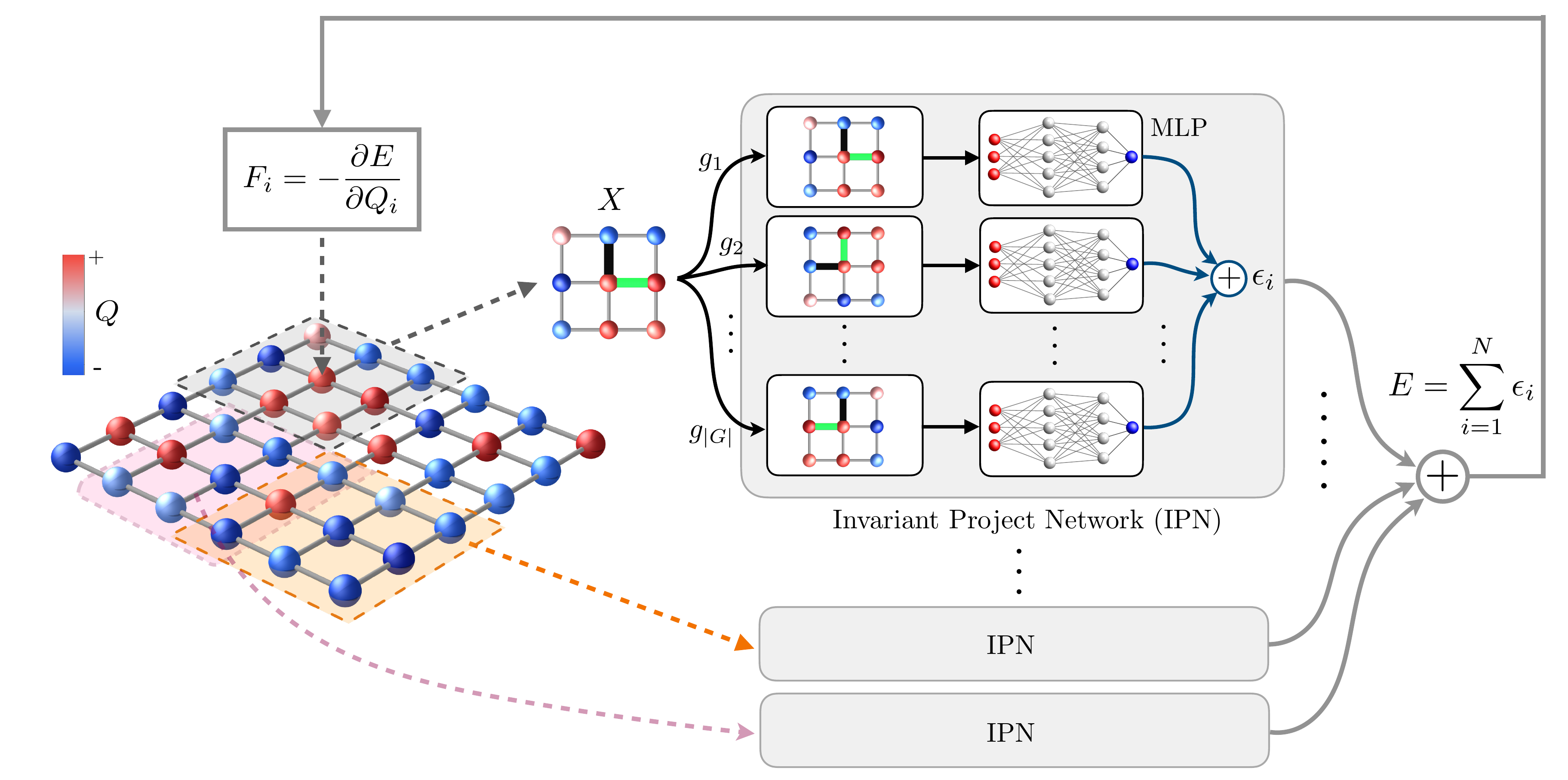}
\caption{Invariant projection network (IPN) for Holstein dynamics. For each lattice site $i$, the local distortion environment $X_i$ is transformed by all $g\in D_4$ and evaluated by a shared MLP; averaging the scalar outputs gives an invariant local energy $\epsilon_i$. Applying the same IPN to overlapping local environments throughout the lattice forms the standard Behler--Parrinello construction, $E_{\rm ML}=\sum_i\epsilon_i$. Differentiating this total energy with respect to the distortions then yields the symmetry-preserving conservative forces $F_i=-\partial E_{\rm ML}/\partial Q_i$.}
\label{fig:Holstein-schematic}
\end{figure*}

Accurate prediction of individual $\Delta E_{i,\alpha}$, however, does not by itself guarantee accurate dynamics. In kMC, even a small error in a transition energy can alter an acceptance decision, after which the ML and ED trajectories generally follow different microscopic paths. A meaningful dynamical benchmark must therefore compare statistical observables rather than individual configurations. We perform EPN-kMC and ED-kMC simulations on a $30\times30$ lattice with $\rho_c=0.55$, $\rho_f=0.187$, $U=2t$, and $T=0.05t$, and monitor the equal-time density correlation $C_f(r)=\langle n_i^f n_j^f\rangle-\rho_f^2$ during relaxation. As shown in Fig.~\ref{fig:FK-benchmark}(b,c), the correlation functions obtained from the two approaches are nearly indistinguishable at both $t=20$ and $100$ sweeps, where one sweep corresponds to $N_f$ attempted hops. The EPN therefore reproduces not only the local transition energetics but also the evolving spatial correlations generated by many successive stochastic updates.

We finally use the EPN to simulate phase ordering on a $150\times150$ square lattice, accessing the same large-scale regime studied in the earlier descriptor-based work~\cite{zhang22b}. Following a thermal quench, the itinerant $c$ electrons mediate effective interactions that drive phase separation and favor local checkerboard charge-density-wave (CDW) order. In a CDW region, the $f$ electrons preferentially occupy one of the two sublattices of the square lattice, producing one of two symmetry-related checkerboard patterns, accompanied by a corresponding modulation of the itinerant-electron density. This microscopic ordering is visible in Fig.~\ref{fig:FK-large}(d), where small checkerboard clusters have already formed. Because the $f$-electron filling is low, however, these CDW regions remain largely disconnected and are separated by metallic regions with few or no $f$ electrons. The ordering on this microscopic scale therefore consists of many spatially separated charge-ordered clusters rather than a single connected checkerboard domain.

A second, much larger length scale nevertheless emerges because these disconnected clusters interact through the itinerant electrons. Each checkerboard cluster carries a sublattice polarity, according to whether its $f$ electrons predominantly occupy the $A$ or $B$ sublattice, and spatially separated clusters show a strong tendency to align this polarity. This produces a hidden $Z_2$ symmetry breaking in addition to the local CDW order. To visualize it, the coarse-grained configurations in Fig.~\ref{fig:FK-large}(a--c) assign each site an Ising variable according to whether the nearest $f$ electron belongs to the $A$ or $B$ sublattice; the red and blue regions therefore represent domains of opposite sublattice polarity, not individual connected CDW clusters. Each such ``supercluster'' can contain many disconnected checkerboard clusters sharing the same polarity. Their progressive growth in Fig.~\ref{fig:FK-large}(a--c) shows that the EPN reproduces this hidden, electron-mediated coarsening. Comparing panels (c) and (d) makes the separation between the microscopic CDW scale and the much larger $Z_2$ ordering scale particularly clear.

\section{Holstein dynamics and conservative forces}
\label{sec:Holstein}

The second application considers the scalar-output limit of the EPN construction for continuous lattice variables. This case connects naturally to the Behler--Parrinello (BP) decomposition~\cite{behler07,behler11}, in which an extensive energy is expressed as a sum of local contributions,
\begin{equation}
E_{\rm ML}=\sum_i \epsilon_i .
\label{eq:BP-energy}
\end{equation}
This basic construction also underlies many modern graph-based and equivariant force fields~\cite{batzner2022,batatia2022}. Setting $D_Y(g)=1$ in Eq.~(\ref{eq:EPN}) gives an invariant projection network (IPN), the scalar special case of the EPN. As illustrated in Fig.~\ref{fig:Holstein-schematic}, the local distortion environment $X_i$ centered at site $i$ is transformed by every $g\in D_4$, and all symmetry-related configurations are evaluated by the same scalar MLP. The resulting local energy is
\begin{equation}
\epsilon_i=\frac{1}{|G|}\sum_{g\in G}f_\theta\!\left[D_X(g)X_i\right],
\label{eq:IPN-energy}
\end{equation}
where $\theta$ denotes the shared MLP parameters. The same parameter-sharing IPN is then applied to the overlapping neighborhoods centered on all lattice sites, and Eq.~(\ref{eq:BP-energy}) assembles these local contributions into the global energy.

The force field follows by differentiating this global potential,
\begin{equation}
F_i^{\rm ML}=-\frac{\partial E_{\rm ML}}{\partial Q_i}
=-\sum_{j:\,i\in\mathcal N_j}\frac{\partial\epsilon_j}{\partial Q_i},
\label{eq:IPN-force}
\end{equation}
where $\mathcal N_j$ is the neighborhood centered on site $j$. A given displacement $Q_i$ belongs to several overlapping neighborhoods, so its force receives contributions from every corresponding local energy. Automatic differentiation through the full BP sum accounts for all of these terms. The same local map is used at every site, and a lattice symmetry permutes the neighborhoods while leaving each scalar prediction invariant. The summed energy is therefore invariant, its gradient has the required transformation law, and the force is conservative by construction. Exact lattice symmetry and the BP energy decomposition are thus combined without first constructing invariant descriptors.

We demonstrate this construction using the semiclassical Holstein model,
\begin{equation}
\begin{aligned}
\mathcal{H}={}&-t\sum_{\langle ij\rangle}\left(c_i^\dagger c_j+\mathrm{H.c.}\right)
-g\sum_i Q_i(n_i-\bar n) \\
&+\sum_i\frac{P_i^2}{2M}
+\frac{K}{2}\sum_i Q_i^2
+\kappa\sum_{\langle ij\rangle}Q_iQ_j ,
\end{aligned}
\label{eq:Holstein}
\end{equation}
where $Q_i$ and $P_i$ are the classical lattice displacement and momentum, $n_i=c_i^\dagger c_i$, $K$ is the local elastic constant, and $\kappa$ couples neighboring distortions. At half filling, $\bar n=1/2$, the low-temperature state develops checkerboard charge-density-wave (CDW) order. Descriptor-based MLP and graph-neural-network force fields have previously been used to simulate its adiabatic dynamics~\cite{cheng23a,Fan2026b}.

As in the FK application, the electrons are assumed to equilibrate rapidly for each instantaneous configuration of the slow variables. Here these variables are the continuous distortions $\{Q_i\}$, and the conservative lattice force derives from the electronic free energy together with the elastic potential at fixed electronic temperature and filling. The ED force labels include both the electronic and elastic contributions, so $E_{\rm ML}$ approximates this full effective potential through its derivatives. Langevin evolution supplements the learned conservative force with damping and thermal noise, as specified in Appendix~\ref{app:Hol-details}. The IPN construction therefore guarantees a conservative force field, while energy exchange with the bath enters through these additional terms.

For the calculations below, each IPN takes a $5\times5$ displacement patch as input. The shared scalar MLP contains hidden layers of widths $512$, $256$, and $128$ with SiLU activations, followed by a linear local-energy output. The same network parameters are used both for all eight symmetry-related configurations entering each projection and for all lattice sites in the BP sum. Training is performed directly against the total conservative forces evaluated using ED for the electronic contribution and the analytic elastic terms. We use 8000 complete $30\times30$ configurations for training and 1000 for testing, with AdamW optimization and a normalized force mean-square loss. The learning rate is annealed from $10^{-4}$ to $10^{-6}$. Force training does not require assigning a unique target energy to each local neighborhood; it constrains the derivatives of the summed energy directly, while the invariant BP construction guarantees that the learned forces derive from a common potential. Details of the dimensionless Langevin dynamics, model parameters, loss function, and optimization are given in Appendix~\ref{app:Hol-details}.

\begin{figure}[t]
    \centering
    \includegraphics[width=0.99\columnwidth]{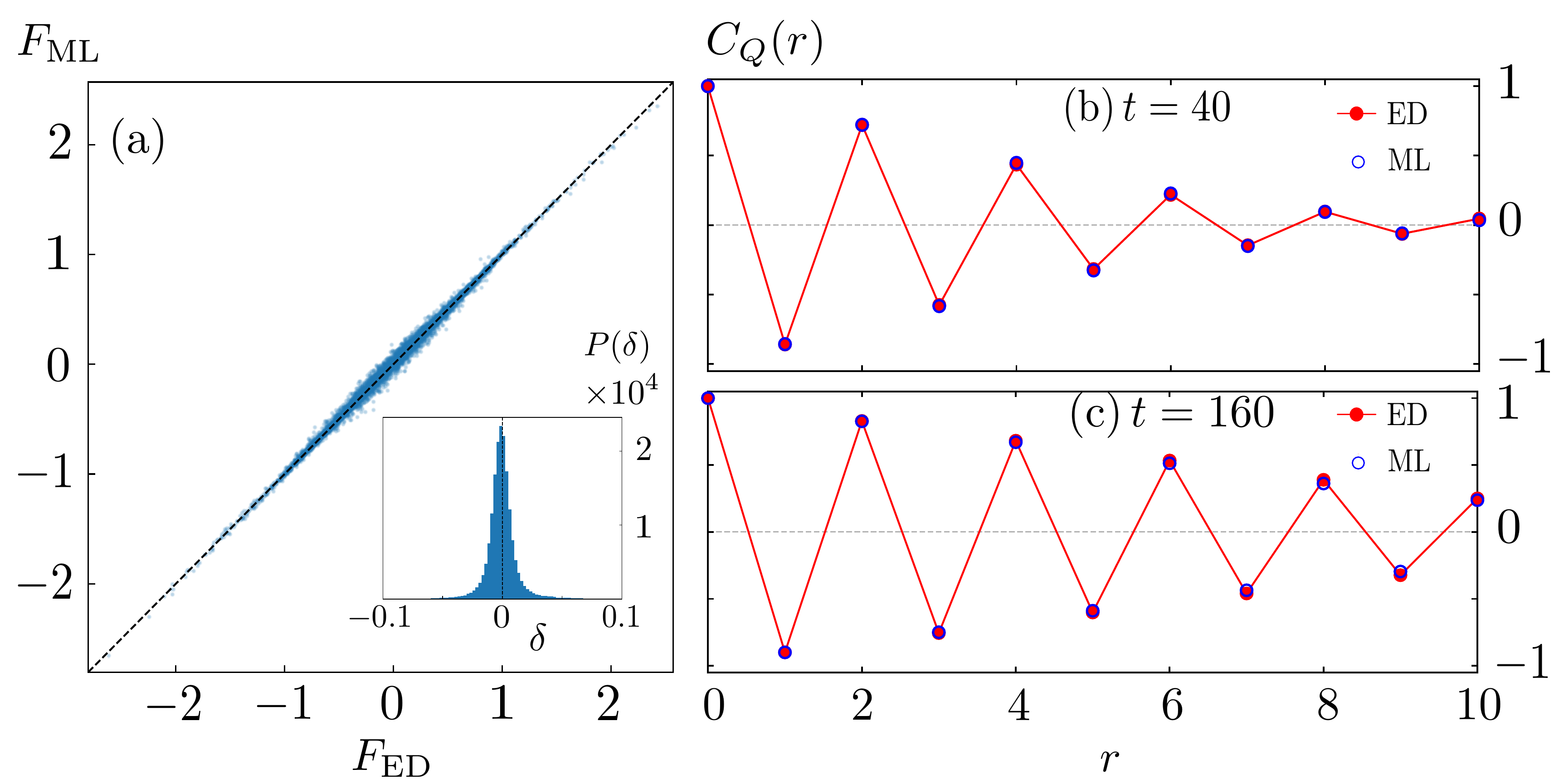}
\caption{Microscopic and dynamical benchmarks of the IPN for the Holstein model. (a) IPN-predicted forces $F_{\rm ML}$ versus exact-diagonalization values $F_{\rm ED}$. The dashed line denotes perfect agreement; the inset shows the distribution of prediction errors $\delta=F_{\rm ML}-F_{\rm ED}$. (b,c) Equal-time lattice-distortion correlation function $C_Q(r)$ obtained from ED and IPN Langevin dynamics after a thermal quench at (b) $t=40$ and (c) $t=160$.}
\label{fig:hol-benchmark}
\end{figure}

Figure~\ref{fig:hol-benchmark}(a) compares the IPN-predicted forces with the ED results over the full test range. The predictions closely follow the diagonal, while the inset shows an error distribution sharply centered near zero. These forces are obtained by differentiating the summed invariant energy, so the agreement tests both the learned local-energy representation and the contributions from overlapping neighborhoods. The benchmark demonstrates that this conservative construction can accurately reproduce the total forces, including their electronic and elastic contributions, directly from the microscopic distortion configurations.

\begin{figure}[t]
    \centering
    \includegraphics[width=0.99\columnwidth]{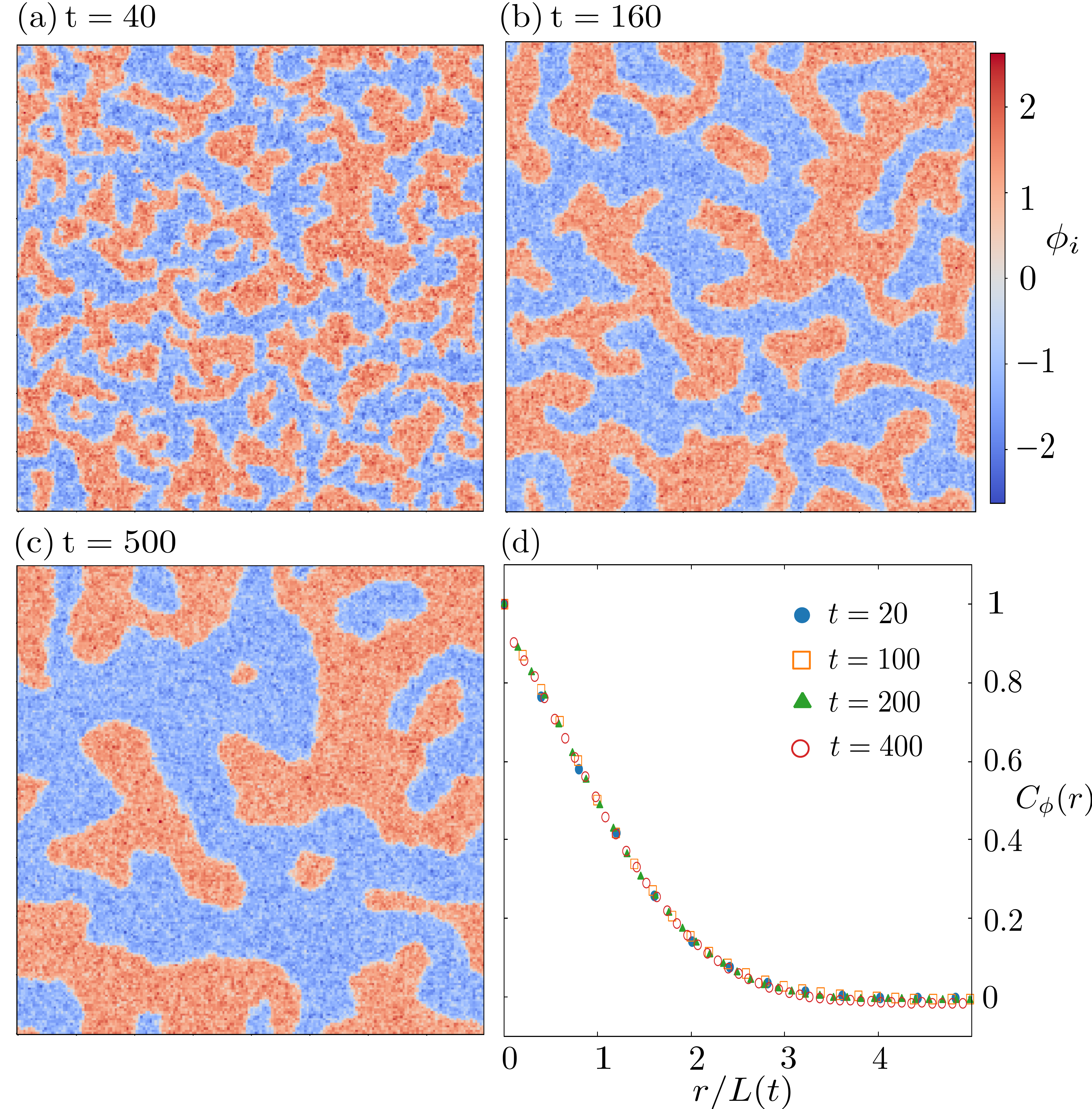}
\caption{Large-scale phase ordering in the Holstein model after a thermal quench on a $200\times200$ square lattice with periodic boundary conditions. (a--c) Snapshots of the CDW order-parameter field $\phi_i$ at $t=40$, $160$, and $500$, showing progressive domain coarsening. (d) Normalized connected order-parameter correlations $C_\phi(r,t)$ plotted against $r/L(t)$, using the power-law growth $L(t)\sim t^{0.454}$. The close collapse of curves at different times supports dynamical scaling.}
\label{fig:hol-large}
\end{figure}

We next embed the learned force in the same Langevin dynamics used for the ED reference. Figures~\ref{fig:hol-benchmark}(b) and \ref{fig:hol-benchmark}(c) compare the equal-time connected distortion correlation $C_Q(r)=\langle Q_iQ_j\rangle-\langle Q\rangle^2$, where $r=|\mathbf r_i-\mathbf r_j|$. The mean displacement vanishes for the checkerboard pattern, while the alternating sign of $C_Q(r)$ reflects the opposite displacements on the two sublattices. The ED and IPN curves remain essentially coincident at both $t=40$ and $160$, reproducing the spatial correlations at different stages of relaxation. This agreement provides a dynamical test beyond the accuracy of individual force predictions: local errors are repeatedly fed back into the evolving displacement field, yet their accumulated effects do not produce appreciable differences in the correlations over the times examined.

Having established agreement with ED on the reference system, we apply the IPN to a thermal quench on a $200\times200$ square lattice with periodic boundary conditions. To visualize the resulting domain structure, we follow Ref.~\cite{cheng23a} and define the local CDW order parameter $\phi_i=(\rho_i-\frac{1}{4}\sum_{j\in{\rm NN}(i)}\rho_j)(-1)^{x_i+y_i}$, where $\rho_i=\langle n_i\rangle$ is the local electron density and ${\rm NN}(i)$ denotes the four nearest neighbors of site $i$. This field measures the local charge modulation, with its sign distinguishing the two checkerboard states related by the broken $Z_2$ sublattice symmetry. The snapshots in Fig.~\ref{fig:hol-large}(a--c), taken at $t=40$, $160$, and $500$, show the progressive coarsening of these domains. Numerous small regions of opposite sign give way to larger domains, making the growth of the characteristic spatial scale apparent.

We quantify this evolution using the normalized connected correlation $C_\phi(r,t)=[\langle\phi_i\phi_j\rangle-\langle\phi\rangle^2]/[\langle\phi^2\rangle-\langle\phi\rangle^2]$, with all averages evaluated at time $t$. The characteristic domain length is extracted from its half-height, $C_\phi(L(t),t)=1/2$. Over the time window studied, this length is well described by a power law $L(t)\sim t^\alpha$ with $\alpha\simeq0.454$. Using this power-law length scale, the correlation curves at different times collapse closely onto a common profile when plotted against $r/L(t)$, as shown in Fig.~\ref{fig:hol-large}(d). Thus, although the domains grow substantially during relaxation, their spatial correlations retain approximately the same form after rescaling by a single evolving length.

The observed exponent is smaller than the conventional Allen--Cahn value $1/2$, consistent with the slower CDW domain growth reported in earlier descriptor-based MLP and graph-based force-field studies~\cite{cheng23a,Fan2026b}. Here $\alpha\simeq0.454$ characterizes the accessible simulation window; the correlation collapse supports a scaling description within this window without establishing the asymptotic growth law. For the present methodological benchmark, the central result is that the IPN reproduces both the microscopic forces and the collective ordering dynamics using an ordinary MLP with invariant projection. Together with the FK results, these tests demonstrate the applicability of the same finite-group construction to directional transition energies and conservative forces for continuous lattice variables.

\section{Discussion and conclusions}
\label{sec:discussion}

In conclusion, finite-group projection provides a direct route from ordinary neural networks to symmetry-preserving surrogates for lattice many-body dynamics. The FK and Holstein applications illustrate complementary cases: directional transition free energies transforming in a nontrivial output representation, and invariant local energies whose derivatives generate conservative forces. In both cases, agreement with exact diagonalization tests the microscopic predictions, while the dynamical correlations and recovery of established coarsening behavior test their accumulated consequences in large-scale simulations. These benchmarks show that accurate dynamics can be obtained without handcrafted invariant descriptors or symmetry-constrained hidden representations.

This is useful because the standard alternatives each carry their own implementation burden. Descriptor-based approaches can be highly effective, but symmetry alone does not determine which invariants, relative phases, reference frames, truncations, or canonical orientations should be used; directional outputs can require still more bookkeeping to restore the orientation information removed by invariant encoding. Fully equivariant neural networks provide a more systematic and often more elegant solution, but implementing them properly generally requires substantially more architectural machinery and group-theoretical structure, including the organization and coupling of symmetry representations throughout the network.

The main advantage of the EPN approach is therefore its simplicity. The microscopic configuration remains the input, an otherwise unrestricted network learns the local map, and a fixed group projection imposes the required transformation law exactly. Changing from scalar to directional, vector, or tensor outputs requires the corresponding output dimension and group representation, while the hidden layers can remain unrestricted. The cost of evaluating the full group average makes this approach especially natural for fixed-lattice problems with small finite symmetry groups; it can also surround more expressive backbones when needed. For many lattice applications, EPNs thus offer a simple, transparent, and reusable route to exact symmetry preservation.

\appendix

\section{Falicov--Kimball simulation and training details}
\label{app:FK-details}

For the FK calculations, a configuration $\mathcal C=\{n_i^f\}$ defines a quadratic itinerant-electron problem. We retain the main-text notation $\Delta E$ for the target, although it is an electronic free-energy difference. For an allowed nearest-neighbor hop $i\rightarrow j$,
\begin{equation}
\Delta E_{i\rightarrow j}=F_e(\mathcal C^{i\rightarrow j})-F_e(\mathcal C),
\end{equation}
with the itinerant-electron number held fixed in the two ED calculations. The kMC acceptance probability is
\begin{equation}
P(\mathcal C\rightarrow\mathcal C^{i\rightarrow j})=\min\!\left[1,\exp\!\left(-\frac{\Delta E_{i\rightarrow j}}{T}\right)\right].
\end{equation}
One sweep contains $N_f$ attempted moves, including rejected proposals. The reference parameters used for the ED and EPN benchmarks are listed in Table~\ref{tab:FK-params}.

\begin{table}[t]
\caption{Falicov--Kimball simulation parameters. Energies and temperatures are measured in units of the nearest-neighbor hopping $t$.}
\label{tab:FK-params}
\begin{ruledtabular}
\begin{tabular}{lc}
Parameter & Value \\
\hline
Lattice dimensions $L_x\times L_y$ & $30\times30$ \\
Itinerant-electron density $\rho_c$ & $0.55$ \\
$f$-electron density $\rho_f$ & $0.187$ \\
Interaction strength $U/t$ & $2.0$ \\
Temperature $T/t$ & $0.05$ \\
Neighborhood radius $R_c$ & $10$ \\
\end{tabular}
\end{ruledtabular}
\end{table}

The local input is the occupation pattern in the radius-$10$ disk centered on the selected occupied site. Including the center, this gives 317 binary variables. The unrestricted MLP is
\begin{equation}
317\rightarrow512\rightarrow512\rightarrow256\rightarrow128\rightarrow4,
\end{equation}
with SiLU activations after all hidden layers and a linear output layer. The four outputs are ordered as $(+x,-x,+y,-y)$. The $D_4$ projection in Eq.~(\ref{eq:EPN}) uses the same MLP parameters for all eight group elements; symmetry therefore changes neither the number of trainable parameters nor the network architecture.

For a training environment $a$ centered on site $i_a$, the mask $m_{a\alpha}=1-n^f_{i_a+e_\alpha}$ removes hops into occupied sites. The loss is the masked normalized mean-square error
\begin{equation}
\mathcal L_{\rm FK}=\frac{\displaystyle\sum_{a,\alpha}m_{a\alpha}\left(\frac{\Delta E^{\rm ML}_{a\alpha}-\Delta E^{\rm ED}_{a\alpha}}{\sigma^{\rm train}_{\Delta E}}\right)^2}{\displaystyle\sum_{a,\alpha}m_{a\alpha}},
\label{eq:FK-loss-app}
\end{equation}
where $\sigma^{\rm train}_{\Delta E}$ is the standard deviation of the valid ED labels in the training set and is held fixed during training and evaluation. Forbidden moves neither contribute to the loss nor enter accepted kMC transitions.

\begin{table}[t]
\caption{FK neural-network and optimization details.}
\label{tab:FK-training}
\begin{ruledtabular}
\begin{tabular}{lc}
Property & Value \\
\hline
Hidden layers & $512,512,256,128$ \\
Activation & SiLU \\
Optimizer & AdamW \\
Learning rate & $10^{-3}$ \\
Weight decay & $10^{-6}$ \\
Batch size & 1024 local environments \\
Training environments & 77,520 \\
Validation environments & 19,380 \\
Training epochs & 500 \\
Training normalized MSE & $1.767\times10^{-3}$ \\
Test normalized MSE & $5.089\times10^{-3}$ \\
\end{tabular}
\end{ruledtabular}
\end{table}

The reported test MSE corresponds to an RMSE of approximately $0.0713\,\sigma^{\rm train}_{\Delta E}$. This is an aggregate error normalized by the spread of the training labels, rather than a relative error for each individual hop.

\section{Holstein simulation and training details}
\label{app:Hol-details}

For the semiclassical Holstein calculations we use the full Hamiltonian
\begin{align}
\mathcal H_{\rm Holstein}={}&-t\sum_{\langle ij\rangle}(c_i^\dagger c_j+\mathrm{H.c.})-g\sum_iQ_i(\hat n_i-\bar n) \nonumber\\
&+\sum_i\frac{P_i^2}{2M}+\frac{K}{2}\sum_iQ_i^2+\kappa\sum_{\langle ij\rangle}Q_iQ_j .
\label{eq:Hol-full-app}
\end{align}
For each displacement configuration, the electronic Hamiltonian is diagonalized and the chemical potential is adjusted to maintain the prescribed filling. In units $E_0=g^2/K$ and $t_0=\sqrt{M/K}$, we define $\widetilde Q_i=KQ_i/g$, $\widetilde t=t_{\rm phys}/t_0$, and $\widetilde P_i=d\widetilde Q_i/d\widetilde t$, together with $\widetilde\kappa=\kappa/K$, $\widetilde T=k_BT/E_0$, and $\widetilde\gamma=\gamma/\sqrt{MK}$. The Langevin dynamics used in the benchmark can be written
\begin{align}
\frac{d\widetilde Q_i}{d\widetilde t}&=\widetilde P_i,\\
\frac{d\widetilde P_i}{d\widetilde t}&=(n_i-\bar n)-\widetilde Q_i-\widetilde\kappa\sum_{j\in {\rm NN}(i)}\widetilde Q_j-\widetilde\gamma\widetilde P_i+\widetilde\xi_i,
\end{align}
with $\langle\widetilde\xi_i(\widetilde t)\widetilde\xi_j(\widetilde t')\rangle=2\widetilde\gamma\widetilde T\,\delta_{ij}\delta(\widetilde t-\widetilde t')$. The parameters are summarized in Table~\ref{tab:Hol-params}.

\begin{table}[t]
\caption{Holstein simulation parameters in the dimensionless convention used for the Langevin dynamics.}
\label{tab:Hol-params}
\begin{ruledtabular}
\begin{tabular}{lc}
Parameter & Value \\
\hline
Lattice dimensions $L_x\times L_y$ & $30\times30$ \\
Filling $\bar n$ & $0.5$ \\
Electron--phonon coupling $\lambda=g^2/(4Kt)$ & $2.0$ \\
Elastic coupling $\widetilde\kappa=\kappa/K$ & $0.15$ \\
Bath temperature $\widetilde T$ & $0.1$ \\
Initial lattice temperature $\widetilde T_{\rm init}$ & $1.0$ \\
Damping $\widetilde\gamma$ & $0.64$ \\
Time step $\Delta\widetilde t$ & $0.02$ \\
Number of time steps & 10,000 \\
\end{tabular}
\end{ruledtabular}
\end{table}

\begin{table}[t]
\caption{Holstein neural-network and optimization details.}
\label{tab:Hol-training}
\begin{ruledtabular}
\begin{tabular}{lc}
Property & Value \\
\hline
Hidden layers & $512,256,128$ \\
Activation & SiLU \\
Optimizer & AdamW \\
Initial learning rate & $10^{-4}$ \\
Final learning rate & $10^{-6}$ \\
Learning-rate schedule & Cosine annealing \\
Weight decay & $10^{-6}$ \\
Batch size & 24 configurations \\
Gradient-norm cutoff & 10 \\
Training configurations & 8000 \\
Test configurations & 1000 \\
Training normalized MSE & $1.021\times10^{-3}$ \\
Test normalized MSE & $1.142\times10^{-3}$ \\
\end{tabular}
\end{ruledtabular}
\end{table}

The scalar MLP takes the 25 displacements in a $5\times5$ patch and has architecture
\begin{equation}
25\rightarrow512\rightarrow256\rightarrow128\rightarrow1,
\end{equation}
again with SiLU hidden activations and a linear output. After invariant $D_4$ averaging, its local energies are summed according to
\begin{equation}
E_{\rm ML}=\sum_j\epsilon_j,\qquad F_i^{\rm ML}=-\frac{\partial E_{\rm ML}}{\partial \widetilde Q_i}.
\end{equation}
Automatic differentiation of the full sum includes contributions from all overlapping patches that contain site $i$. The ED force labels include both electronic and elastic contributions, so the learned force represents the full conservative force entering the Langevin equation. Damping and stochastic noise are added separately during the dynamics.

Training minimizes the normalized force mean-square error
\begin{equation}
\mathcal L_{\rm Holstein}=\frac{1}{N_{\rm conf}N}\sum_{a,i}\left(\frac{F^{\rm ML}_{ai}-F^{\rm ED}_{ai}}{F^{\rm train}_{\rm rms}}\right)^2,
\label{eq:Hol-loss-app}
\end{equation}
where $N=900$ for the $30\times30$ reference lattice and
\begin{equation}
F^{\rm train}_{\rm rms}=\left[\frac{1}{N_{\rm train}N}\sum_{a\in {\rm train},i}(F^{\rm ED}_{ai})^2\right]^{1/2}.
\end{equation}
This normalization is determined from the training data and then kept fixed for both optimization and test evaluation.

The Holstein test MSE corresponds to an RMSE of approximately $0.0338\,F^{\rm train}_{\rm rms}$. Because the FK and Holstein losses use different physical targets and normalization scales, their numerical loss values should not be compared directly.

\bibliography{ref}

\begin{thebibliography}{63}%
\makeatletter
\providecommand \@ifxundefined [1]{%
 \@ifx{#1\undefined}
}%
\providecommand \@ifnum [1]{%
 \ifnum #1\expandafter \@firstoftwo
 \else \expandafter \@secondoftwo
 \fi
}%
\providecommand \@ifx [1]{%
 \ifx #1\expandafter \@firstoftwo
 \else \expandafter \@secondoftwo
 \fi
}%
\providecommand \natexlab [1]{#1}%
\providecommand \enquote  [1]{``#1''}%
\providecommand \bibnamefont  [1]{#1}%
\providecommand \bibfnamefont [1]{#1}%
\providecommand \citenamefont [1]{#1}%
\providecommand \href@noop [0]{\@secondoftwo}%
\providecommand \href [0]{\begingroup \@sanitize@url \@href}%
\providecommand \@href[1]{\@@startlink{#1}\@@href}%
\providecommand \@@href[1]{\endgroup#1\@@endlink}%
\providecommand \@sanitize@url [0]{\catcode `\\12\catcode `\$12\catcode
  `\&12\catcode `\#12\catcode `\^12\catcode `\_12\catcode `\%12\relax}%
\providecommand \@@startlink[1]{}%
\providecommand \@@endlink[0]{}%
\providecommand \url  [0]{\begingroup\@sanitize@url \@url }%
\providecommand \@url [1]{\endgroup\@href {#1}{\urlprefix }}%
\providecommand \urlprefix  [0]{URL }%
\providecommand \Eprint [0]{\href }%
\providecommand \doibase [0]{https://doi.org/}%
\providecommand \selectlanguage [0]{\@gobble}%
\providecommand \bibinfo  [0]{\@secondoftwo}%
\providecommand \bibfield  [0]{\@secondoftwo}%
\providecommand \translation [1]{[#1]}%
\providecommand \BibitemOpen [0]{}%
\providecommand \bibitemStop [0]{}%
\providecommand \bibitemNoStop [0]{.\EOS\space}%
\providecommand \EOS [0]{\spacefactor3000\relax}%
\providecommand \BibitemShut  [1]{\csname bibitem#1\endcsname}%
\let\auto@bib@innerbib\@empty
%</preamble>
\bibitem [{\citenamefont {Dagotto}(2005)}]{dagotto05}%
  \BibitemOpen
  \bibfield  {author} {\bibinfo {author} {\bibfnamefont {E.}~\bibnamefont
  {Dagotto}},\ }\bibfield  {title} {\bibinfo {title} {Complexity in strongly
  correlated electronic systems},\ }\href
  {https://doi.org/10.1126/science.1107559} {\bibfield  {journal} {\bibinfo
  {journal} {Science}\ }\textbf {\bibinfo {volume} {309}},\ \bibinfo {pages}
  {257} (\bibinfo {year} {2005})}\BibitemShut {NoStop}%
\bibitem [{\citenamefont {Kivelson}\ \emph {et~al.}(1998)\citenamefont
  {Kivelson}, \citenamefont {Fradkin},\ and\ \citenamefont
  {Emery}}]{kivelson98}%
  \BibitemOpen
  \bibfield  {author} {\bibinfo {author} {\bibfnamefont {S.~A.}\ \bibnamefont
  {Kivelson}}, \bibinfo {author} {\bibfnamefont {E.}~\bibnamefont {Fradkin}},\
  and\ \bibinfo {author} {\bibfnamefont {V.~J.}\ \bibnamefont {Emery}},\
  }\bibfield  {title} {\bibinfo {title} {Electronic liquid-crystal phases of a
  doped {Mott} insulator},\ }\href {https://doi.org/10.1038/31177} {\bibfield
  {journal} {\bibinfo  {journal} {Nature}\ }\textbf {\bibinfo {volume} {393}},\
  \bibinfo {pages} {550} (\bibinfo {year} {1998})}\BibitemShut {NoStop}%
\bibitem [{\citenamefont {Fradkin}\ \emph {et~al.}(2015)\citenamefont
  {Fradkin}, \citenamefont {Kivelson},\ and\ \citenamefont
  {Tranquada}}]{fradkin15}%
  \BibitemOpen
  \bibfield  {author} {\bibinfo {author} {\bibfnamefont {E.}~\bibnamefont
  {Fradkin}}, \bibinfo {author} {\bibfnamefont {S.~A.}\ \bibnamefont
  {Kivelson}},\ and\ \bibinfo {author} {\bibfnamefont {J.~M.}\ \bibnamefont
  {Tranquada}},\ }\bibfield  {title} {\bibinfo {title} {Colloquium: Theory of
  intertwined orders in high temperature superconductors},\ }\href
  {https://doi.org/10.1103/RevModPhys.87.457} {\bibfield  {journal} {\bibinfo
  {journal} {Rev. Mod. Phys.}\ }\textbf {\bibinfo {volume} {87}},\ \bibinfo
  {pages} {457} (\bibinfo {year} {2015})}\BibitemShut {NoStop}%
\bibitem [{\citenamefont {Hamidian}\ \emph {et~al.}(2016)\citenamefont
  {Hamidian}, \citenamefont {Edkins}, \citenamefont {Joo}, \citenamefont
  {Kostin}, \citenamefont {Eisaki}, \citenamefont {Uchida}, \citenamefont
  {Lawler}, \citenamefont {Kim}, \citenamefont {Mackenzie}, \citenamefont
  {Fujita}, \citenamefont {Lee},\ and\ \citenamefont {Davis}}]{hamidian16}%
  \BibitemOpen
  \bibfield  {author} {\bibinfo {author} {\bibfnamefont {M.~H.}\ \bibnamefont
  {Hamidian}}, \bibinfo {author} {\bibfnamefont {S.~D.}\ \bibnamefont
  {Edkins}}, \bibinfo {author} {\bibfnamefont {S.~H.}\ \bibnamefont {Joo}},
  \bibinfo {author} {\bibfnamefont {A.}~\bibnamefont {Kostin}}, \bibinfo
  {author} {\bibfnamefont {H.}~\bibnamefont {Eisaki}}, \bibinfo {author}
  {\bibfnamefont {S.}~\bibnamefont {Uchida}}, \bibinfo {author} {\bibfnamefont
  {M.~J.}\ \bibnamefont {Lawler}}, \bibinfo {author} {\bibfnamefont {E.-A.}\
  \bibnamefont {Kim}}, \bibinfo {author} {\bibfnamefont {A.~P.}\ \bibnamefont
  {Mackenzie}}, \bibinfo {author} {\bibfnamefont {K.}~\bibnamefont {Fujita}},
  \bibinfo {author} {\bibfnamefont {J.}~\bibnamefont {Lee}},\ and\ \bibinfo
  {author} {\bibfnamefont {J.~C.~S.}\ \bibnamefont {Davis}},\ }\bibfield
  {title} {\bibinfo {title} {Detection of a cooper-pair density wave in
  {Bi$_2$Sr$_2$CaCu$_2$O$_{8+x}$}},\ }\href
  {https://doi.org/10.1038/nature17411} {\bibfield  {journal} {\bibinfo
  {journal} {Nature}\ }\textbf {\bibinfo {volume} {532}},\ \bibinfo {pages}
  {343} (\bibinfo {year} {2016})}\BibitemShut {NoStop}%
\bibitem [{\citenamefont {Qazilbash}\ \emph {et~al.}(2007)\citenamefont
  {Qazilbash}, \citenamefont {Brehm}, \citenamefont {Chae}, \citenamefont {Ho},
  \citenamefont {Andreev}, \citenamefont {Kim}, \citenamefont {Yun},
  \citenamefont {Balatsky}, \citenamefont {Maple}, \citenamefont {Keilmann},
  \citenamefont {Kim},\ and\ \citenamefont {Basov}}]{qazilbash07}%
  \BibitemOpen
  \bibfield  {author} {\bibinfo {author} {\bibfnamefont {M.~M.}\ \bibnamefont
  {Qazilbash}}, \bibinfo {author} {\bibfnamefont {M.}~\bibnamefont {Brehm}},
  \bibinfo {author} {\bibfnamefont {B.-G.}\ \bibnamefont {Chae}}, \bibinfo
  {author} {\bibfnamefont {P.-C.}\ \bibnamefont {Ho}}, \bibinfo {author}
  {\bibfnamefont {G.~O.}\ \bibnamefont {Andreev}}, \bibinfo {author}
  {\bibfnamefont {B.-J.}\ \bibnamefont {Kim}}, \bibinfo {author} {\bibfnamefont
  {S.~J.}\ \bibnamefont {Yun}}, \bibinfo {author} {\bibfnamefont {A.~V.}\
  \bibnamefont {Balatsky}}, \bibinfo {author} {\bibfnamefont {M.~B.}\
  \bibnamefont {Maple}}, \bibinfo {author} {\bibfnamefont {F.}~\bibnamefont
  {Keilmann}}, \bibinfo {author} {\bibfnamefont {H.-T.}\ \bibnamefont {Kim}},\
  and\ \bibinfo {author} {\bibfnamefont {D.~N.}\ \bibnamefont {Basov}},\
  }\bibfield  {title} {\bibinfo {title} {Mott transition in {VO$_2$} revealed
  by infrared spectroscopy and nano-imaging},\ }\href
  {https://doi.org/10.1126/science.1150124} {\bibfield  {journal} {\bibinfo
  {journal} {Science}\ }\textbf {\bibinfo {volume} {318}},\ \bibinfo {pages}
  {1750} (\bibinfo {year} {2007})}\BibitemShut {NoStop}%
\bibitem [{\citenamefont {McLeod}\ \emph {et~al.}(2017)\citenamefont {McLeod},
  \citenamefont {van Heumen}, \citenamefont {Ramirez}, \citenamefont {Wang},
  \citenamefont {Saerbeck}, \citenamefont {Guenon}, \citenamefont {Goldflam},
  \citenamefont {Anderegg}, \citenamefont {Kelly}, \citenamefont {Mueller},
  \citenamefont {Liu}, \citenamefont {Schuller},\ and\ \citenamefont
  {Basov}}]{mcleod17}%
  \BibitemOpen
  \bibfield  {author} {\bibinfo {author} {\bibfnamefont {A.~S.}\ \bibnamefont
  {McLeod}}, \bibinfo {author} {\bibfnamefont {E.}~\bibnamefont {van Heumen}},
  \bibinfo {author} {\bibfnamefont {J.~G.}\ \bibnamefont {Ramirez}}, \bibinfo
  {author} {\bibfnamefont {S.}~\bibnamefont {Wang}}, \bibinfo {author}
  {\bibfnamefont {T.}~\bibnamefont {Saerbeck}}, \bibinfo {author}
  {\bibfnamefont {S.}~\bibnamefont {Guenon}}, \bibinfo {author} {\bibfnamefont
  {M.}~\bibnamefont {Goldflam}}, \bibinfo {author} {\bibfnamefont
  {L.}~\bibnamefont {Anderegg}}, \bibinfo {author} {\bibfnamefont
  {P.}~\bibnamefont {Kelly}}, \bibinfo {author} {\bibfnamefont
  {A.}~\bibnamefont {Mueller}}, \bibinfo {author} {\bibfnamefont {M.~K.}\
  \bibnamefont {Liu}}, \bibinfo {author} {\bibfnamefont {I.~K.}\ \bibnamefont
  {Schuller}},\ and\ \bibinfo {author} {\bibfnamefont {D.~N.}\ \bibnamefont
  {Basov}},\ }\bibfield  {title} {\bibinfo {title} {Nanotextured phase
  coexistence in the correlated insulator {V$_2$O$_3$}},\ }\href
  {https://doi.org/10.1038/nphys3882} {\bibfield  {journal} {\bibinfo
  {journal} {Nature Physics}\ }\textbf {\bibinfo {volume} {13}},\ \bibinfo
  {pages} {80} (\bibinfo {year} {2017})}\BibitemShut {NoStop}%
\bibitem [{\citenamefont {Tokura}\ \emph {et~al.}(2017)\citenamefont {Tokura},
  \citenamefont {Kawasaki},\ and\ \citenamefont {Nagaosa}}]{Tokura2017}%
  \BibitemOpen
  \bibfield  {author} {\bibinfo {author} {\bibfnamefont {Y.}~\bibnamefont
  {Tokura}}, \bibinfo {author} {\bibfnamefont {M.}~\bibnamefont {Kawasaki}},\
  and\ \bibinfo {author} {\bibfnamefont {N.}~\bibnamefont {Nagaosa}},\
  }\bibfield  {title} {\bibinfo {title} {Emergent functions of quantum
  materials},\ }\href {https://doi.org/10.1038/nphys4274} {\bibfield  {journal}
  {\bibinfo  {journal} {Nature Physics}\ }\textbf {\bibinfo {volume} {13}},\
  \bibinfo {pages} {1056} (\bibinfo {year} {2017})}\BibitemShut {NoStop}%
\bibitem [{\citenamefont {Nagaosa}\ and\ \citenamefont
  {Tokura}(2013)}]{nagaosa13}%
  \BibitemOpen
  \bibfield  {author} {\bibinfo {author} {\bibfnamefont {N.}~\bibnamefont
  {Nagaosa}}\ and\ \bibinfo {author} {\bibfnamefont {Y.}~\bibnamefont
  {Tokura}},\ }\bibfield  {title} {\bibinfo {title} {Topological properties and
  dynamics of magnetic skyrmions},\ }\href
  {https://doi.org/10.1038/nnano.2013.243} {\bibfield  {journal} {\bibinfo
  {journal} {Nature Nanotechnology}\ }\textbf {\bibinfo {volume} {8}},\
  \bibinfo {pages} {899} (\bibinfo {year} {2013})}\BibitemShut {NoStop}%
\bibitem [{\citenamefont {Kurumaji}\ \emph {et~al.}(2019)\citenamefont
  {Kurumaji}, \citenamefont {Nakajima}, \citenamefont {Hirschberger},
  \citenamefont {Kikkawa}, \citenamefont {Yamasaki}, \citenamefont {Sagayama},
  \citenamefont {Nakao}, \citenamefont {Taguchi}, \citenamefont {Arima},\ and\
  \citenamefont {Tokura}}]{Kurumaji2019}%
  \BibitemOpen
  \bibfield  {author} {\bibinfo {author} {\bibfnamefont {T.}~\bibnamefont
  {Kurumaji}}, \bibinfo {author} {\bibfnamefont {T.}~\bibnamefont {Nakajima}},
  \bibinfo {author} {\bibfnamefont {M.}~\bibnamefont {Hirschberger}}, \bibinfo
  {author} {\bibfnamefont {A.}~\bibnamefont {Kikkawa}}, \bibinfo {author}
  {\bibfnamefont {Y.}~\bibnamefont {Yamasaki}}, \bibinfo {author}
  {\bibfnamefont {H.}~\bibnamefont {Sagayama}}, \bibinfo {author}
  {\bibfnamefont {H.}~\bibnamefont {Nakao}}, \bibinfo {author} {\bibfnamefont
  {Y.}~\bibnamefont {Taguchi}}, \bibinfo {author} {\bibfnamefont {T.-h.}\
  \bibnamefont {Arima}},\ and\ \bibinfo {author} {\bibfnamefont
  {Y.}~\bibnamefont {Tokura}},\ }\bibfield  {title} {\bibinfo {title} {Skyrmion
  lattice with a giant topological hall effect in a frustrated
  triangular-lattice magnet},\ }\href {https://doi.org/10.1126/science.aau0968}
  {\bibfield  {journal} {\bibinfo  {journal} {Science}\ }\textbf {\bibinfo
  {volume} {365}},\ \bibinfo {pages} {914} (\bibinfo {year}
  {2019})}\BibitemShut {NoStop}%
\bibitem [{\citenamefont {Hohenberg}\ and\ \citenamefont
  {Halperin}(1977)}]{hohenberg77}%
  \BibitemOpen
  \bibfield  {author} {\bibinfo {author} {\bibfnamefont {P.~C.}\ \bibnamefont
  {Hohenberg}}\ and\ \bibinfo {author} {\bibfnamefont {B.~I.}\ \bibnamefont
  {Halperin}},\ }\bibfield  {title} {\bibinfo {title} {Theory of dynamic
  critical phenomena},\ }\href {https://doi.org/10.1103/RevModPhys.49.435}
  {\bibfield  {journal} {\bibinfo  {journal} {Rev. Mod. Phys.}\ }\textbf
  {\bibinfo {volume} {49}},\ \bibinfo {pages} {435} (\bibinfo {year}
  {1977})}\BibitemShut {NoStop}%
\bibitem [{\citenamefont {Bray}(1994)}]{bray94}%
  \BibitemOpen
  \bibfield  {author} {\bibinfo {author} {\bibfnamefont {A.~J.}\ \bibnamefont
  {Bray}},\ }\bibfield  {title} {\bibinfo {title} {Theory of phase-ordering
  kinetics},\ }\href {https://doi.org/10.1080/00018739400101505} {\bibfield
  {journal} {\bibinfo  {journal} {Advances in Physics}\ }\textbf {\bibinfo
  {volume} {43}},\ \bibinfo {pages} {357} (\bibinfo {year} {1994})}\BibitemShut
  {NoStop}%
\bibitem [{\citenamefont {Furukawa}(1985)}]{furukawa85}%
  \BibitemOpen
  \bibfield  {author} {\bibinfo {author} {\bibfnamefont {H.}~\bibnamefont
  {Furukawa}},\ }\bibfield  {title} {\bibinfo {title} {A dynamic scaling
  assumption for phase separation},\ }\href
  {https://doi.org/10.1080/00018738500101841} {\bibfield  {journal} {\bibinfo
  {journal} {Advances in Physics}\ }\textbf {\bibinfo {volume} {34}},\ \bibinfo
  {pages} {703} (\bibinfo {year} {1985})}\BibitemShut {NoStop}%
\bibitem [{\citenamefont {Behler}\ and\ \citenamefont
  {Parrinello}(2007)}]{behler07}%
  \BibitemOpen
  \bibfield  {author} {\bibinfo {author} {\bibfnamefont {J.}~\bibnamefont
  {Behler}}\ and\ \bibinfo {author} {\bibfnamefont {M.}~\bibnamefont
  {Parrinello}},\ }\bibfield  {title} {\bibinfo {title} {Generalized
  neural-network representation of high-dimensional potential-energy
  surfaces},\ }\href {https://doi.org/10.1103/PhysRevLett.98.146401} {\bibfield
   {journal} {\bibinfo  {journal} {Phys. Rev. Lett.}\ }\textbf {\bibinfo
  {volume} {98}},\ \bibinfo {pages} {146401} (\bibinfo {year}
  {2007})}\BibitemShut {NoStop}%
\bibitem [{\citenamefont {Bart\'ok}\ \emph {et~al.}(2010)\citenamefont
  {Bart\'ok}, \citenamefont {Payne}, \citenamefont {Kondor},\ and\
  \citenamefont {Cs\'anyi}}]{bartok10}%
  \BibitemOpen
  \bibfield  {author} {\bibinfo {author} {\bibfnamefont {A.~P.}\ \bibnamefont
  {Bart\'ok}}, \bibinfo {author} {\bibfnamefont {M.~C.}\ \bibnamefont {Payne}},
  \bibinfo {author} {\bibfnamefont {R.}~\bibnamefont {Kondor}},\ and\ \bibinfo
  {author} {\bibfnamefont {G.}~\bibnamefont {Cs\'anyi}},\ }\bibfield  {title}
  {\bibinfo {title} {Gaussian approximation potentials: The accuracy of quantum
  mechanics, without the electrons},\ }\href
  {https://doi.org/10.1103/PhysRevLett.104.136403} {\bibfield  {journal}
  {\bibinfo  {journal} {Phys. Rev. Lett.}\ }\textbf {\bibinfo {volume} {104}},\
  \bibinfo {pages} {136403} (\bibinfo {year} {2010})}\BibitemShut {NoStop}%
\bibitem [{\citenamefont {Li}\ \emph {et~al.}(2015)\citenamefont {Li},
  \citenamefont {Kermode},\ and\ \citenamefont {De~Vita}}]{li15}%
  \BibitemOpen
  \bibfield  {author} {\bibinfo {author} {\bibfnamefont {Z.}~\bibnamefont
  {Li}}, \bibinfo {author} {\bibfnamefont {J.~R.}\ \bibnamefont {Kermode}},\
  and\ \bibinfo {author} {\bibfnamefont {A.}~\bibnamefont {De~Vita}},\
  }\bibfield  {title} {\bibinfo {title} {Molecular dynamics with on-the-fly
  machine learning of quantum-mechanical forces},\ }\href
  {https://doi.org/10.1103/PhysRevLett.114.096405} {\bibfield  {journal}
  {\bibinfo  {journal} {Phys. Rev. Lett.}\ }\textbf {\bibinfo {volume} {114}},\
  \bibinfo {pages} {096405} (\bibinfo {year} {2015})}\BibitemShut {NoStop}%
\bibitem [{\citenamefont {Smith}\ \emph {et~al.}(2017)\citenamefont {Smith},
  \citenamefont {Isayev},\ and\ \citenamefont {Roitberg}}]{smith17}%
  \BibitemOpen
  \bibfield  {author} {\bibinfo {author} {\bibfnamefont {J.~S.}\ \bibnamefont
  {Smith}}, \bibinfo {author} {\bibfnamefont {O.}~\bibnamefont {Isayev}},\ and\
  \bibinfo {author} {\bibfnamefont {A.~E.}\ \bibnamefont {Roitberg}},\
  }\bibfield  {title} {\bibinfo {title} {Ani-1: an extensible neural network
  potential with dft accuracy at force field computational cost},\ }\href
  {https://doi.org/10.1039/C6SC05720A} {\bibfield  {journal} {\bibinfo
  {journal} {Chem. Sci.}\ }\textbf {\bibinfo {volume} {8}},\ \bibinfo {pages}
  {3192} (\bibinfo {year} {2017})}\BibitemShut {NoStop}%
\bibitem [{\citenamefont {Chmiela}\ \emph {et~al.}(2017)\citenamefont
  {Chmiela}, \citenamefont {Tkatchenko}, \citenamefont {Sauceda}, \citenamefont
  {Poltavsky}, \citenamefont {Schütt},\ and\ \citenamefont
  {Müller}}]{chmiela17}%
  \BibitemOpen
  \bibfield  {author} {\bibinfo {author} {\bibfnamefont {S.}~\bibnamefont
  {Chmiela}}, \bibinfo {author} {\bibfnamefont {A.}~\bibnamefont {Tkatchenko}},
  \bibinfo {author} {\bibfnamefont {H.~E.}\ \bibnamefont {Sauceda}}, \bibinfo
  {author} {\bibfnamefont {I.}~\bibnamefont {Poltavsky}}, \bibinfo {author}
  {\bibfnamefont {K.~T.}\ \bibnamefont {Schütt}},\ and\ \bibinfo {author}
  {\bibfnamefont {K.-R.}\ \bibnamefont {Müller}},\ }\bibfield  {title}
  {\bibinfo {title} {Machine learning of accurate energy-conserving molecular
  force fields},\ }\href {https://doi.org/10.1126/sciadv.1603015} {\bibfield
  {journal} {\bibinfo  {journal} {Science Advances}\ }\textbf {\bibinfo
  {volume} {3}},\ \bibinfo {pages} {e1603015} (\bibinfo {year}
  {2017})}\BibitemShut {NoStop}%
\bibitem [{\citenamefont {Zhang}\ \emph {et~al.}(2018)\citenamefont {Zhang},
  \citenamefont {Han}, \citenamefont {Wang}, \citenamefont {Car},\ and\
  \citenamefont {E}}]{zhang18}%
  \BibitemOpen
  \bibfield  {author} {\bibinfo {author} {\bibfnamefont {L.}~\bibnamefont
  {Zhang}}, \bibinfo {author} {\bibfnamefont {J.}~\bibnamefont {Han}}, \bibinfo
  {author} {\bibfnamefont {H.}~\bibnamefont {Wang}}, \bibinfo {author}
  {\bibfnamefont {R.}~\bibnamefont {Car}},\ and\ \bibinfo {author}
  {\bibfnamefont {W.}~\bibnamefont {E}},\ }\bibfield  {title} {\bibinfo {title}
  {Deep potential molecular dynamics: A scalable model with the accuracy of
  quantum mechanics},\ }\href {https://doi.org/10.1103/PhysRevLett.120.143001}
  {\bibfield  {journal} {\bibinfo  {journal} {Phys. Rev. Lett.}\ }\textbf
  {\bibinfo {volume} {120}},\ \bibinfo {pages} {143001} (\bibinfo {year}
  {2018})}\BibitemShut {NoStop}%
\bibitem [{\citenamefont {Deringer}\ \emph {et~al.}(2019)\citenamefont
  {Deringer}, \citenamefont {Caro},\ and\ \citenamefont {Csanyi}}]{deringer19}%
  \BibitemOpen
  \bibfield  {author} {\bibinfo {author} {\bibfnamefont {V.~L.}\ \bibnamefont
  {Deringer}}, \bibinfo {author} {\bibfnamefont {M.~A.}\ \bibnamefont {Caro}},\
  and\ \bibinfo {author} {\bibfnamefont {G.}~\bibnamefont {Csanyi}},\
  }\bibfield  {title} {\bibinfo {title} {Machine learning interatomic
  potentials as emerging tools for materials science},\ }\href
  {https://doi.org/10.1002/adma.201902765} {\bibfield  {journal} {\bibinfo
  {journal} {Advanced Materials}\ }\textbf {\bibinfo {volume} {31}},\ \bibinfo
  {pages} {1902765} (\bibinfo {year} {2019})}\BibitemShut {NoStop}%
\bibitem [{\citenamefont {No{\'e}}\ \emph {et~al.}(2020)\citenamefont
  {No{\'e}}, \citenamefont {Tkatchenko}, \citenamefont {M{\"u}ller},\ and\
  \citenamefont {Clementi}}]{noe20}%
  \BibitemOpen
  \bibfield  {author} {\bibinfo {author} {\bibfnamefont {F.}~\bibnamefont
  {No{\'e}}}, \bibinfo {author} {\bibfnamefont {A.}~\bibnamefont {Tkatchenko}},
  \bibinfo {author} {\bibfnamefont {K.-R.}\ \bibnamefont {M{\"u}ller}},\ and\
  \bibinfo {author} {\bibfnamefont {C.}~\bibnamefont {Clementi}},\ }\bibfield
  {title} {\bibinfo {title} {Machine learning for molecular simulation},\
  }\href {https://doi.org/10.1146/annurev-physchem-042018-052331} {\bibfield
  {journal} {\bibinfo  {journal} {Annu. Rev. Phys. Chem.}\ }\textbf {\bibinfo
  {volume} {71}},\ \bibinfo {pages} {361} (\bibinfo {year} {2020})}\BibitemShut
  {NoStop}%
\bibitem [{\citenamefont {Suwa}\ \emph {et~al.}(2019)\citenamefont {Suwa},
  \citenamefont {Smith}, \citenamefont {Lubbers}, \citenamefont {Batista},
  \citenamefont {Chern},\ and\ \citenamefont {Barros}}]{suwa19}%
  \BibitemOpen
  \bibfield  {author} {\bibinfo {author} {\bibfnamefont {H.}~\bibnamefont
  {Suwa}}, \bibinfo {author} {\bibfnamefont {J.~S.}\ \bibnamefont {Smith}},
  \bibinfo {author} {\bibfnamefont {N.}~\bibnamefont {Lubbers}}, \bibinfo
  {author} {\bibfnamefont {C.~D.}\ \bibnamefont {Batista}}, \bibinfo {author}
  {\bibfnamefont {G.-W.}\ \bibnamefont {Chern}},\ and\ \bibinfo {author}
  {\bibfnamefont {K.}~\bibnamefont {Barros}},\ }\bibfield  {title} {\bibinfo
  {title} {Machine learning for molecular dynamics with strongly correlated
  electrons},\ }\href {https://doi.org/10.1103/PhysRevB.99.161107} {\bibfield
  {journal} {\bibinfo  {journal} {Phys. Rev. B}\ }\textbf {\bibinfo {volume}
  {99}},\ \bibinfo {pages} {161107} (\bibinfo {year} {2019})}\BibitemShut
  {NoStop}%
\bibitem [{\citenamefont {Zhang}\ and\ \citenamefont {Chern}(2021)}]{zhang21}%
  \BibitemOpen
  \bibfield  {author} {\bibinfo {author} {\bibfnamefont {P.}~\bibnamefont
  {Zhang}}\ and\ \bibinfo {author} {\bibfnamefont {G.-W.}\ \bibnamefont
  {Chern}},\ }\bibfield  {title} {\bibinfo {title} {Arrested phase separation
  in double-exchange models: Large-scale simulation enabled by machine
  learning},\ }\href {https://doi.org/10.1103/PhysRevLett.127.146401}
  {\bibfield  {journal} {\bibinfo  {journal} {Phys. Rev. Lett.}\ }\textbf
  {\bibinfo {volume} {127}},\ \bibinfo {pages} {146401} (\bibinfo {year}
  {2021})}\BibitemShut {NoStop}%
\bibitem [{\citenamefont {Zhang}\ \emph
  {et~al.}(2022{\natexlab{a}})\citenamefont {Zhang}, \citenamefont {Zhang},\
  and\ \citenamefont {Chern}}]{zhang22b}%
  \BibitemOpen
  \bibfield  {author} {\bibinfo {author} {\bibfnamefont {S.}~\bibnamefont
  {Zhang}}, \bibinfo {author} {\bibfnamefont {P.}~\bibnamefont {Zhang}},\ and\
  \bibinfo {author} {\bibfnamefont {G.-W.}\ \bibnamefont {Chern}},\ }\bibfield
  {title} {\bibinfo {title} {Anomalous phase separation in a correlated
  electron system: Machine-learning enabled large-scale kinetic monte carlo
  simulations},\ }\href {https://doi.org/10.1073/pnas.2119957119} {\bibfield
  {journal} {\bibinfo  {journal} {Proceedings of the National Academy of
  Sciences}\ }\textbf {\bibinfo {volume} {119}},\ \bibinfo {pages}
  {e2119957119} (\bibinfo {year} {2022}{\natexlab{a}})}\BibitemShut {NoStop}%
\bibitem [{\citenamefont {Cheng}\ \emph
  {et~al.}(2023{\natexlab{a}})\citenamefont {Cheng}, \citenamefont {Zhang},\
  and\ \citenamefont {Chern}}]{cheng23a}%
  \BibitemOpen
  \bibfield  {author} {\bibinfo {author} {\bibfnamefont {C.}~\bibnamefont
  {Cheng}}, \bibinfo {author} {\bibfnamefont {S.}~\bibnamefont {Zhang}},\ and\
  \bibinfo {author} {\bibfnamefont {G.-W.}\ \bibnamefont {Chern}},\ }\bibfield
  {title} {\bibinfo {title} {Machine learning for phase ordering dynamics of
  charge density waves},\ }\href {https://doi.org/10.1103/PhysRevB.108.014301}
  {\bibfield  {journal} {\bibinfo  {journal} {Phys. Rev. B}\ }\textbf {\bibinfo
  {volume} {108}},\ \bibinfo {pages} {014301} (\bibinfo {year}
  {2023}{\natexlab{a}})}\BibitemShut {NoStop}%
\bibitem [{\citenamefont {Ghosh}\ \emph {et~al.}(2024)\citenamefont {Ghosh},
  \citenamefont {Zhang}, \citenamefont {Cheng},\ and\ \citenamefont
  {Chern}}]{Ghosh24}%
  \BibitemOpen
  \bibfield  {author} {\bibinfo {author} {\bibfnamefont {S.}~\bibnamefont
  {Ghosh}}, \bibinfo {author} {\bibfnamefont {S.}~\bibnamefont {Zhang}},
  \bibinfo {author} {\bibfnamefont {C.}~\bibnamefont {Cheng}},\ and\ \bibinfo
  {author} {\bibfnamefont {G.-W.}\ \bibnamefont {Chern}},\ }\bibfield  {title}
  {\bibinfo {title} {Kinetics of orbital ordering in cooperative jahn-teller
  models: Machine-learning enabled large-scale simulations},\ }\href
  {https://doi.org/10.1103/PhysRevMaterials.8.123602} {\bibfield  {journal}
  {\bibinfo  {journal} {Phys. Rev. Mater.}\ }\textbf {\bibinfo {volume} {8}},\
  \bibinfo {pages} {123602} (\bibinfo {year} {2024})}\BibitemShut {NoStop}%
\bibitem [{\citenamefont {Cheng}\ \emph
  {et~al.}(2023{\natexlab{b}})\citenamefont {Cheng}, \citenamefont {Zhang},
  \citenamefont {Nguyen}, \citenamefont {Azarfar}, \citenamefont {Chern},\ and\
  \citenamefont {Baek}}]{cheng23b}%
  \BibitemOpen
  \bibfield  {author} {\bibinfo {author} {\bibfnamefont {X.}~\bibnamefont
  {Cheng}}, \bibinfo {author} {\bibfnamefont {S.}~\bibnamefont {Zhang}},
  \bibinfo {author} {\bibfnamefont {P.~C.~H.}\ \bibnamefont {Nguyen}}, \bibinfo
  {author} {\bibfnamefont {S.}~\bibnamefont {Azarfar}}, \bibinfo {author}
  {\bibfnamefont {G.-W.}\ \bibnamefont {Chern}},\ and\ \bibinfo {author}
  {\bibfnamefont {S.~S.}\ \bibnamefont {Baek}},\ }\bibfield  {title} {\bibinfo
  {title} {Convolutional neural networks for large-scale dynamical modeling of
  itinerant magnets},\ }\href
  {https://doi.org/10.1103/PhysRevResearch.5.033188} {\bibfield  {journal}
  {\bibinfo  {journal} {Phys. Rev. Res.}\ }\textbf {\bibinfo {volume} {5}},\
  \bibinfo {pages} {033188} (\bibinfo {year} {2023}{\natexlab{b}})}\BibitemShut
  {NoStop}%
\bibitem [{\citenamefont {Zhang}\ and\ \citenamefont {Chern}(2023)}]{zhang23}%
  \BibitemOpen
  \bibfield  {author} {\bibinfo {author} {\bibfnamefont {P.}~\bibnamefont
  {Zhang}}\ and\ \bibinfo {author} {\bibfnamefont {G.-W.}\ \bibnamefont
  {Chern}},\ }\bibfield  {title} {\bibinfo {title} {Machine learning
  nonequilibrium electron forces for spin dynamics of itinerant magnets},\
  }\href {https://doi.org/10.1038/s41524-023-00990-0} {\bibfield  {journal}
  {\bibinfo  {journal} {npj Computational Materials}\ }\textbf {\bibinfo
  {volume} {9}},\ \bibinfo {pages} {32} (\bibinfo {year} {2023})}\BibitemShut
  {NoStop}%
\bibitem [{\citenamefont {Fan}\ \emph {et~al.}(2024)\citenamefont {Fan},
  \citenamefont {Zhang},\ and\ \citenamefont {Chern}}]{Fan24}%
  \BibitemOpen
  \bibfield  {author} {\bibinfo {author} {\bibfnamefont {Y.}~\bibnamefont
  {Fan}}, \bibinfo {author} {\bibfnamefont {S.}~\bibnamefont {Zhang}},\ and\
  \bibinfo {author} {\bibfnamefont {G.-W.}\ \bibnamefont {Chern}},\ }\bibfield
  {title} {\bibinfo {title} {Coarsening of chiral domains in itinerant electron
  magnets: A machine learning force-field approach},\ }\href
  {https://doi.org/10.1103/PhysRevB.110.245105} {\bibfield  {journal} {\bibinfo
   {journal} {Phys. Rev. B}\ }\textbf {\bibinfo {volume} {110}},\ \bibinfo
  {pages} {245105} (\bibinfo {year} {2024})}\BibitemShut {NoStop}%
\bibitem [{\citenamefont {Tyberg}\ \emph {et~al.}(2025)\citenamefont {Tyberg},
  \citenamefont {Fan},\ and\ \citenamefont {Chern}}]{tyberg25}%
  \BibitemOpen
  \bibfield  {author} {\bibinfo {author} {\bibfnamefont {A.}~\bibnamefont
  {Tyberg}}, \bibinfo {author} {\bibfnamefont {Y.}~\bibnamefont {Fan}},\ and\
  \bibinfo {author} {\bibfnamefont {G.-W.}\ \bibnamefont {Chern}},\ }\bibfield
  {title} {\bibinfo {title} {Machine learning force field model for kinetic
  monte carlo simulations of itinerant ising magnets},\ }\href
  {https://doi.org/10.1103/d3zm-pbr1} {\bibfield  {journal} {\bibinfo
  {journal} {Phys. Rev. B}\ }\textbf {\bibinfo {volume} {111}},\ \bibinfo
  {pages} {235132} (\bibinfo {year} {2025})}\BibitemShut {NoStop}%
\bibitem [{\citenamefont {Chern}\ \emph {et~al.}(2026)\citenamefont {Chern},
  \citenamefont {Fan}, \citenamefont {Zhang},\ and\ \citenamefont
  {Zhang}}]{Chern2026b}%
  \BibitemOpen
  \bibfield  {author} {\bibinfo {author} {\bibfnamefont {G.-W.}\ \bibnamefont
  {Chern}}, \bibinfo {author} {\bibfnamefont {Y.}~\bibnamefont {Fan}}, \bibinfo
  {author} {\bibfnamefont {S.}~\bibnamefont {Zhang}},\ and\ \bibinfo {author}
  {\bibfnamefont {P.}~\bibnamefont {Zhang}},\ }\bibfield  {title} {\bibinfo
  {title} {Machine-learning modeling of magnetization dynamics in itinerant
  magnets},\ }\href {https://doi.org/10.1016/j.jmmm.2026.173679} {\bibfield
  {journal} {\bibinfo  {journal} {Journal of Magnetism and Magnetic Materials}\
  }\textbf {\bibinfo {volume} {628}},\ \bibinfo {pages} {173679} (\bibinfo
  {year} {2026})}\BibitemShut {NoStop}%
\bibitem [{\citenamefont {Behler}(2011)}]{behler11}%
  \BibitemOpen
  \bibfield  {author} {\bibinfo {author} {\bibfnamefont {J.}~\bibnamefont
  {Behler}},\ }\bibfield  {title} {\bibinfo {title} {{Atom-centered symmetry
  functions for constructing high-dimensional neural network potentials}},\
  }\href {https://doi.org/10.1063/1.3553717} {\bibfield  {journal} {\bibinfo
  {journal} {The Journal of Chemical Physics}\ }\textbf {\bibinfo {volume}
  {134}},\ \bibinfo {pages} {074106} (\bibinfo {year} {2011})}\BibitemShut
  {NoStop}%
\bibitem [{\citenamefont {Bart\'ok}\ \emph {et~al.}(2013)\citenamefont
  {Bart\'ok}, \citenamefont {Kondor},\ and\ \citenamefont
  {Cs\'anyi}}]{bartok13}%
  \BibitemOpen
  \bibfield  {author} {\bibinfo {author} {\bibfnamefont {A.~P.}\ \bibnamefont
  {Bart\'ok}}, \bibinfo {author} {\bibfnamefont {R.}~\bibnamefont {Kondor}},\
  and\ \bibinfo {author} {\bibfnamefont {G.}~\bibnamefont {Cs\'anyi}},\
  }\bibfield  {title} {\bibinfo {title} {On representing chemical
  environments},\ }\href {https://doi.org/10.1103/PhysRevB.87.184115}
  {\bibfield  {journal} {\bibinfo  {journal} {Phys. Rev. B}\ }\textbf {\bibinfo
  {volume} {87}},\ \bibinfo {pages} {184115} (\bibinfo {year}
  {2013})}\BibitemShut {NoStop}%
\bibitem [{\citenamefont {Ghiringhelli}\ \emph {et~al.}(2015)\citenamefont
  {Ghiringhelli}, \citenamefont {Vybiral}, \citenamefont {Levchenko},
  \citenamefont {Draxl},\ and\ \citenamefont {Scheffler}}]{ghiringhelli15}%
  \BibitemOpen
  \bibfield  {author} {\bibinfo {author} {\bibfnamefont {L.~M.}\ \bibnamefont
  {Ghiringhelli}}, \bibinfo {author} {\bibfnamefont {J.}~\bibnamefont
  {Vybiral}}, \bibinfo {author} {\bibfnamefont {S.~V.}\ \bibnamefont
  {Levchenko}}, \bibinfo {author} {\bibfnamefont {C.}~\bibnamefont {Draxl}},\
  and\ \bibinfo {author} {\bibfnamefont {M.}~\bibnamefont {Scheffler}},\
  }\bibfield  {title} {\bibinfo {title} {Big data of materials science:
  Critical role of the descriptor},\ }\href
  {https://doi.org/10.1103/PhysRevLett.114.105503} {\bibfield  {journal}
  {\bibinfo  {journal} {Phys. Rev. Lett.}\ }\textbf {\bibinfo {volume} {114}},\
  \bibinfo {pages} {105503} (\bibinfo {year} {2015})}\BibitemShut {NoStop}%
\bibitem [{\citenamefont {Himanen}\ \emph {et~al.}(2020)\citenamefont
  {Himanen}, \citenamefont {Jäger}, \citenamefont {Morooka}, \citenamefont
  {{Federici Canova}}, \citenamefont {Ranawat}, \citenamefont {Gao},
  \citenamefont {Rinke},\ and\ \citenamefont {Foster}}]{himanen20}%
  \BibitemOpen
  \bibfield  {author} {\bibinfo {author} {\bibfnamefont {L.}~\bibnamefont
  {Himanen}}, \bibinfo {author} {\bibfnamefont {M.~O.}\ \bibnamefont {Jäger}},
  \bibinfo {author} {\bibfnamefont {E.~V.}\ \bibnamefont {Morooka}}, \bibinfo
  {author} {\bibfnamefont {F.}~\bibnamefont {{Federici Canova}}}, \bibinfo
  {author} {\bibfnamefont {Y.~S.}\ \bibnamefont {Ranawat}}, \bibinfo {author}
  {\bibfnamefont {D.~Z.}\ \bibnamefont {Gao}}, \bibinfo {author} {\bibfnamefont
  {P.}~\bibnamefont {Rinke}},\ and\ \bibinfo {author} {\bibfnamefont {A.~S.}\
  \bibnamefont {Foster}},\ }\bibfield  {title} {\bibinfo {title} {Dscribe:
  Library of descriptors for machine learning in materials science},\ }\href
  {https://doi.org/https://doi.org/10.1016/j.cpc.2019.106949} {\bibfield
  {journal} {\bibinfo  {journal} {Computer Physics Communications}\ }\textbf
  {\bibinfo {volume} {247}},\ \bibinfo {pages} {106949} (\bibinfo {year}
  {2020})}\BibitemShut {NoStop}%
\bibitem [{\citenamefont {Huo}\ and\ \citenamefont {Rupp}(2022)}]{huo22}%
  \BibitemOpen
  \bibfield  {author} {\bibinfo {author} {\bibfnamefont {H.}~\bibnamefont
  {Huo}}\ and\ \bibinfo {author} {\bibfnamefont {M.}~\bibnamefont {Rupp}},\
  }\bibfield  {title} {\bibinfo {title} {Unified representation of molecules
  and crystals for machine learning},\ }\href
  {https://doi.org/10.1088/2632-2153/aca005} {\bibfield  {journal} {\bibinfo
  {journal} {Machine Learning: Science and Technology}\ }\textbf {\bibinfo
  {volume} {3}},\ \bibinfo {pages} {045017} (\bibinfo {year}
  {2022})}\BibitemShut {NoStop}%
\bibitem [{\citenamefont {Drautz}(2019)}]{drautz19}%
  \BibitemOpen
  \bibfield  {author} {\bibinfo {author} {\bibfnamefont {R.}~\bibnamefont
  {Drautz}},\ }\bibfield  {title} {\bibinfo {title} {Atomic cluster expansion
  for accurate and transferable interatomic potentials},\ }\href
  {https://doi.org/10.1103/PhysRevB.99.014104} {\bibfield  {journal} {\bibinfo
  {journal} {Phys. Rev. B}\ }\textbf {\bibinfo {volume} {99}},\ \bibinfo
  {pages} {014104} (\bibinfo {year} {2019})}\BibitemShut {NoStop}%
\bibitem [{\citenamefont {Zhang}\ \emph
  {et~al.}(2022{\natexlab{b}})\citenamefont {Zhang}, \citenamefont {Zhang},\
  and\ \citenamefont {Chern}}]{zhang22}%
  \BibitemOpen
  \bibfield  {author} {\bibinfo {author} {\bibfnamefont {P.}~\bibnamefont
  {Zhang}}, \bibinfo {author} {\bibfnamefont {S.}~\bibnamefont {Zhang}},\ and\
  \bibinfo {author} {\bibfnamefont {G.-W.}\ \bibnamefont {Chern}},\ }\href@noop
  {} {\bibinfo {title} {Descriptors for machine learning model of generalized
  force field in condensed matter systems}} (\bibinfo {year}
  {2022}{\natexlab{b}}),\ \Eprint {https://arxiv.org/abs/2201.00798}
  {arXiv:2201.00798 [cond-mat.str-el]} \BibitemShut {NoStop}%
\bibitem [{\citenamefont {Cohen}\ and\ \citenamefont
  {Welling}(2016)}]{cohen2016}%
  \BibitemOpen
  \bibfield  {author} {\bibinfo {author} {\bibfnamefont {T.~S.}\ \bibnamefont
  {Cohen}}\ and\ \bibinfo {author} {\bibfnamefont {M.}~\bibnamefont
  {Welling}},\ }\bibfield  {title} {\bibinfo {title} {Group equivariant
  convolutional networks},\ }in\ \href
  {https://proceedings.mlr.press/v48/cohenc16.html} {\emph {\bibinfo
  {booktitle} {Proceedings of the 33rd International Conference on Machine
  Learning}}},\ \bibinfo {series} {Proceedings of Machine Learning Research},
  Vol.~\bibinfo {volume} {48}\ (\bibinfo  {publisher} {PMLR},\ \bibinfo {year}
  {2016})\ pp.\ \bibinfo {pages} {2990--2999},\ \Eprint
  {https://arxiv.org/abs/1602.07576} {arXiv:1602.07576 [cs.LG]} \BibitemShut
  {NoStop}%
\bibitem [{\citenamefont {Cohen}\ \emph {et~al.}(2018)\citenamefont {Cohen},
  \citenamefont {Geiger}, \citenamefont {Weiler},\ and\ \citenamefont
  {Welling}}]{cohen2018}%
  \BibitemOpen
  \bibfield  {author} {\bibinfo {author} {\bibfnamefont {T.~S.}\ \bibnamefont
  {Cohen}}, \bibinfo {author} {\bibfnamefont {M.}~\bibnamefont {Geiger}},
  \bibinfo {author} {\bibfnamefont {M.}~\bibnamefont {Weiler}},\ and\ \bibinfo
  {author} {\bibfnamefont {M.}~\bibnamefont {Welling}},\ }\bibfield  {title}
  {\bibinfo {title} {A general theory of equivariant cnns on homogeneous
  spaces},\ }in\ \href
  {https://proceedings.neurips.cc/paper/2018/hash/488e4104520c6aab692863cc1dba45af-Abstract.html}
  {\emph {\bibinfo {booktitle} {Advances in Neural Information Processing
  Systems}}},\ Vol.~\bibinfo {volume} {31}\ (\bibinfo {year} {2018})\ \Eprint
  {https://arxiv.org/abs/1811.02017} {arXiv:1811.02017 [cs.LG]} \BibitemShut
  {NoStop}%
\bibitem [{\citenamefont {Thomas}\ \emph {et~al.}(2018)\citenamefont {Thomas},
  \citenamefont {Smidt}, \citenamefont {Kearnes}, \citenamefont {Yang},
  \citenamefont {Li}, \citenamefont {Kohlhoff},\ and\ \citenamefont
  {Riley}}]{thomas2018}%
  \BibitemOpen
  \bibfield  {author} {\bibinfo {author} {\bibfnamefont {N.}~\bibnamefont
  {Thomas}}, \bibinfo {author} {\bibfnamefont {T.}~\bibnamefont {Smidt}},
  \bibinfo {author} {\bibfnamefont {S.}~\bibnamefont {Kearnes}}, \bibinfo
  {author} {\bibfnamefont {L.}~\bibnamefont {Yang}}, \bibinfo {author}
  {\bibfnamefont {L.}~\bibnamefont {Li}}, \bibinfo {author} {\bibfnamefont
  {K.}~\bibnamefont {Kohlhoff}},\ and\ \bibinfo {author} {\bibfnamefont
  {P.}~\bibnamefont {Riley}},\ }\href {https://arxiv.org/abs/1802.08219}
  {\bibinfo {title} {Tensor field networks: Rotation- and
  translation-equivariant neural networks for 3d point clouds}} (\bibinfo
  {year} {2018}),\ \Eprint {https://arxiv.org/abs/1802.08219} {arXiv:1802.08219
  [cs.LG]} \BibitemShut {NoStop}%
\bibitem [{\citenamefont {Weiler}\ \emph {et~al.}(2018)\citenamefont {Weiler},
  \citenamefont {Geiger}, \citenamefont {Welling}, \citenamefont {Boomsma},\
  and\ \citenamefont {Cohen}}]{weiler2018}%
  \BibitemOpen
  \bibfield  {author} {\bibinfo {author} {\bibfnamefont {M.}~\bibnamefont
  {Weiler}}, \bibinfo {author} {\bibfnamefont {M.}~\bibnamefont {Geiger}},
  \bibinfo {author} {\bibfnamefont {M.}~\bibnamefont {Welling}}, \bibinfo
  {author} {\bibfnamefont {W.}~\bibnamefont {Boomsma}},\ and\ \bibinfo {author}
  {\bibfnamefont {T.~S.}\ \bibnamefont {Cohen}},\ }\bibfield  {title} {\bibinfo
  {title} {3d steerable cnns: Learning rotationally equivariant features in
  volumetric data},\ }in\ \href
  {https://proceedings.neurips.cc/paper/2018/hash/488e4104520c6aab692863cc1dba45af-Abstract.html}
  {\emph {\bibinfo {booktitle} {Advances in Neural Information Processing
  Systems}}},\ Vol.~\bibinfo {volume} {31}\ (\bibinfo  {publisher} {Curran
  Associates, Inc.},\ \bibinfo {year} {2018})\ \Eprint
  {https://arxiv.org/abs/1807.02547} {arXiv:1807.02547 [cs.LG]} \BibitemShut
  {NoStop}%
\bibitem [{\citenamefont {Geiger}\ and\ \citenamefont
  {Smidt}(2022)}]{Geiger22}%
  \BibitemOpen
  \bibfield  {author} {\bibinfo {author} {\bibfnamefont {M.}~\bibnamefont
  {Geiger}}\ and\ \bibinfo {author} {\bibfnamefont {T.}~\bibnamefont {Smidt}},\
  }\bibfield  {title} {\bibinfo {title} {e3nn: Euclidean neural networks},\
  }\href {https://arxiv.org/abs/2207.09453} {\bibfield  {journal} {\bibinfo
  {journal} {arXiv preprint}\ } (\bibinfo {year} {2022})},\ \Eprint
  {https://arxiv.org/abs/2207.09453} {arXiv:2207.09453 [cs.LG]} \BibitemShut
  {NoStop}%
\bibitem [{\citenamefont {Batzner}\ \emph {et~al.}(2022)\citenamefont
  {Batzner}, \citenamefont {Musaelian}, \citenamefont {Sun}, \citenamefont
  {Geiger}, \citenamefont {Mailoa}, \citenamefont {Kornbluth}, \citenamefont
  {Molinari}, \citenamefont {Smidt},\ and\ \citenamefont
  {Kozinsky}}]{batzner2022}%
  \BibitemOpen
  \bibfield  {author} {\bibinfo {author} {\bibfnamefont {S.}~\bibnamefont
  {Batzner}}, \bibinfo {author} {\bibfnamefont {A.}~\bibnamefont {Musaelian}},
  \bibinfo {author} {\bibfnamefont {L.}~\bibnamefont {Sun}}, \bibinfo {author}
  {\bibfnamefont {M.}~\bibnamefont {Geiger}}, \bibinfo {author} {\bibfnamefont
  {J.~P.}\ \bibnamefont {Mailoa}}, \bibinfo {author} {\bibfnamefont
  {M.}~\bibnamefont {Kornbluth}}, \bibinfo {author} {\bibfnamefont
  {N.}~\bibnamefont {Molinari}}, \bibinfo {author} {\bibfnamefont {T.~E.}\
  \bibnamefont {Smidt}},\ and\ \bibinfo {author} {\bibfnamefont
  {B.}~\bibnamefont {Kozinsky}},\ }\bibfield  {title} {\bibinfo {title}
  {E(3)-equivariant graph neural networks for data-efficient and accurate
  interatomic potentials},\ }\href {https://doi.org/10.1038/s41467-022-29939-5}
  {\bibfield  {journal} {\bibinfo  {journal} {Nature Communications}\ }\textbf
  {\bibinfo {volume} {13}},\ \bibinfo {pages} {2453} (\bibinfo {year}
  {2022})}\BibitemShut {NoStop}%
\bibitem [{\citenamefont {Batatia}\ \emph {et~al.}(2022)\citenamefont
  {Batatia}, \citenamefont {Kovács}, \citenamefont {Simm}, \citenamefont
  {Ortner},\ and\ \citenamefont {Cs\'anyi}}]{batatia2022}%
  \BibitemOpen
  \bibfield  {author} {\bibinfo {author} {\bibfnamefont {I.}~\bibnamefont
  {Batatia}}, \bibinfo {author} {\bibfnamefont {D.~P.}\ \bibnamefont
  {Kovács}}, \bibinfo {author} {\bibfnamefont {G.~N.~C.}\ \bibnamefont
  {Simm}}, \bibinfo {author} {\bibfnamefont {C.}~\bibnamefont {Ortner}},\ and\
  \bibinfo {author} {\bibfnamefont {G.}~\bibnamefont {Cs\'anyi}},\ }\bibfield
  {title} {\bibinfo {title} {Mace: Higher order equivariant message passing
  neural networks for fast and accurate force fields},\ }in\ \href
  {https://proceedings.neurips.cc/paper/2022/hash/4a36c3c51af11ed9f34615b81edb5bbc-Abstract-Conference.html}
  {\emph {\bibinfo {booktitle} {Advances in Neural Information Processing
  Systems}}},\ Vol.~\bibinfo {volume} {35}\ (\bibinfo {year} {2022})\ pp.\
  \bibinfo {pages} {11423--11436},\ \Eprint {https://arxiv.org/abs/2206.07697}
  {arXiv:2206.07697 [stat.ML]} \BibitemShut {NoStop}%
\bibitem [{\citenamefont {Musaelian}\ \emph {et~al.}(2023)\citenamefont
  {Musaelian}, \citenamefont {Batzner}, \citenamefont {Johansson},
  \citenamefont {Sun}, \citenamefont {Owen}, \citenamefont {Kornbluth},\ and\
  \citenamefont {Kozinsky}}]{musaelian2023}%
  \BibitemOpen
  \bibfield  {author} {\bibinfo {author} {\bibfnamefont {A.}~\bibnamefont
  {Musaelian}}, \bibinfo {author} {\bibfnamefont {S.}~\bibnamefont {Batzner}},
  \bibinfo {author} {\bibfnamefont {A.}~\bibnamefont {Johansson}}, \bibinfo
  {author} {\bibfnamefont {L.}~\bibnamefont {Sun}}, \bibinfo {author}
  {\bibfnamefont {C.~J.}\ \bibnamefont {Owen}}, \bibinfo {author}
  {\bibfnamefont {M.}~\bibnamefont {Kornbluth}},\ and\ \bibinfo {author}
  {\bibfnamefont {B.}~\bibnamefont {Kozinsky}},\ }\bibfield  {title} {\bibinfo
  {title} {Learning local equivariant representations for large-scale atomistic
  dynamics},\ }\href {https://doi.org/10.1038/s41467-023-36329-y} {\bibfield
  {journal} {\bibinfo  {journal} {Nature Communications}\ }\textbf {\bibinfo
  {volume} {14}},\ \bibinfo {pages} {579} (\bibinfo {year} {2023})}\BibitemShut
  {NoStop}%
\bibitem [{\citenamefont {Puny}\ \emph {et~al.}(2022)\citenamefont {Puny},
  \citenamefont {Atzmon}, \citenamefont {Ben-Hamu}, \citenamefont {Misra},
  \citenamefont {Grover}, \citenamefont {Smith},\ and\ \citenamefont
  {Lipman}}]{puny2022}%
  \BibitemOpen
  \bibfield  {author} {\bibinfo {author} {\bibfnamefont {O.}~\bibnamefont
  {Puny}}, \bibinfo {author} {\bibfnamefont {M.}~\bibnamefont {Atzmon}},
  \bibinfo {author} {\bibfnamefont {H.}~\bibnamefont {Ben-Hamu}}, \bibinfo
  {author} {\bibfnamefont {I.}~\bibnamefont {Misra}}, \bibinfo {author}
  {\bibfnamefont {A.}~\bibnamefont {Grover}}, \bibinfo {author} {\bibfnamefont
  {E.~J.}\ \bibnamefont {Smith}},\ and\ \bibinfo {author} {\bibfnamefont
  {Y.}~\bibnamefont {Lipman}},\ }\bibfield  {title} {\bibinfo {title} {Frame
  averaging for invariant and equivariant network design},\ }in\ \href
  {https://openreview.net/forum?id=zIUyj55nXR} {\emph {\bibinfo {booktitle}
  {International Conference on Learning Representations}}}\ (\bibinfo {year}
  {2022})\ \Eprint {https://arxiv.org/abs/2110.03336} {arXiv:2110.03336
  [cs.LG]} \BibitemShut {NoStop}%
\bibitem [{\citenamefont {Basu}\ \emph {et~al.}(2023)\citenamefont {Basu},
  \citenamefont {Sattigeri}, \citenamefont {Natesan~Ramamurthy}, \citenamefont
  {Chenthamarakshan}, \citenamefont {Varshney}, \citenamefont {Varshney},\ and\
  \citenamefont {Das}}]{basu2023a}%
  \BibitemOpen
  \bibfield  {author} {\bibinfo {author} {\bibfnamefont {S.}~\bibnamefont
  {Basu}}, \bibinfo {author} {\bibfnamefont {P.}~\bibnamefont {Sattigeri}},
  \bibinfo {author} {\bibfnamefont {K.}~\bibnamefont {Natesan~Ramamurthy}},
  \bibinfo {author} {\bibfnamefont {V.}~\bibnamefont {Chenthamarakshan}},
  \bibinfo {author} {\bibfnamefont {K.~R.}\ \bibnamefont {Varshney}}, \bibinfo
  {author} {\bibfnamefont {L.~R.}\ \bibnamefont {Varshney}},\ and\ \bibinfo
  {author} {\bibfnamefont {P.}~\bibnamefont {Das}},\ }\bibfield  {title}
  {\bibinfo {title} {{Equi-Tuning}: Group equivariant fine-tuning of pretrained
  models},\ }\href {https://doi.org/10.1609/aaai.v37i6.25832} {\bibfield
  {journal} {\bibinfo  {journal} {Proceedings of the AAAI Conference on
  Artificial Intelligence}\ }\textbf {\bibinfo {volume} {37}},\ \bibinfo
  {pages} {6788} (\bibinfo {year} {2023})},\ \Eprint
  {https://arxiv.org/abs/2210.06475} {arXiv:2210.06475 [cs.LG]} \BibitemShut
  {NoStop}%
\bibitem [{\citenamefont {Sannai}\ \emph {et~al.}(2024)\citenamefont {Sannai},
  \citenamefont {Kawano},\ and\ \citenamefont {Kumagai}}]{sannai2024}%
  \BibitemOpen
  \bibfield  {author} {\bibinfo {author} {\bibfnamefont {A.}~\bibnamefont
  {Sannai}}, \bibinfo {author} {\bibfnamefont {M.}~\bibnamefont {Kawano}},\
  and\ \bibinfo {author} {\bibfnamefont {W.}~\bibnamefont {Kumagai}},\
  }\bibfield  {title} {\bibinfo {title} {Invariant and equivariant {Reynolds}
  networks},\ }\href {https://www.jmlr.org/papers/v25/22-0891.html} {\bibfield
  {journal} {\bibinfo  {journal} {Journal of Machine Learning Research}\
  }\textbf {\bibinfo {volume} {25}},\ \bibinfo {pages} {1} (\bibinfo {year}
  {2024})},\ \Eprint {https://arxiv.org/abs/2110.08092} {arXiv:2110.08092
  [cs.LG]} \BibitemShut {NoStop}%
\bibitem [{\citenamefont {Dittmer}\ \emph {et~al.}(2022)\citenamefont
  {Dittmer}, \citenamefont {Erzmann}, \citenamefont {Harms},\ and\
  \citenamefont {Maass}}]{dittmer2022}%
  \BibitemOpen
  \bibfield  {author} {\bibinfo {author} {\bibfnamefont {S.}~\bibnamefont
  {Dittmer}}, \bibinfo {author} {\bibfnamefont {D.}~\bibnamefont {Erzmann}},
  \bibinfo {author} {\bibfnamefont {H.}~\bibnamefont {Harms}},\ and\ \bibinfo
  {author} {\bibfnamefont {P.}~\bibnamefont {Maass}},\ }\href
  {https://doi.org/10.48550/arXiv.2209.05098} {\bibinfo {title} {{SELTO}:
  Sample-efficient learned topology optimization}} (\bibinfo {year} {2022}),\
  \Eprint {https://arxiv.org/abs/2209.05098} {arXiv:2209.05098 [cs.LG]}
  \BibitemShut {NoStop}%
\bibitem [{\citenamefont {Corbetta}\ \emph {et~al.}(2023)\citenamefont
  {Corbetta}, \citenamefont {Gabbana}, \citenamefont {Gyrya}, \citenamefont
  {Livescu}, \citenamefont {Prins},\ and\ \citenamefont
  {Toschi}}]{corbetta2023}%
  \BibitemOpen
  \bibfield  {author} {\bibinfo {author} {\bibfnamefont {A.}~\bibnamefont
  {Corbetta}}, \bibinfo {author} {\bibfnamefont {A.}~\bibnamefont {Gabbana}},
  \bibinfo {author} {\bibfnamefont {V.}~\bibnamefont {Gyrya}}, \bibinfo
  {author} {\bibfnamefont {D.}~\bibnamefont {Livescu}}, \bibinfo {author}
  {\bibfnamefont {J.}~\bibnamefont {Prins}},\ and\ \bibinfo {author}
  {\bibfnamefont {F.}~\bibnamefont {Toschi}},\ }\bibfield  {title} {\bibinfo
  {title} {Toward learning {Lattice Boltzmann} collision operators},\ }\href
  {https://doi.org/10.1140/epje/s10189-023-00267-w} {\bibfield  {journal}
  {\bibinfo  {journal} {Eur. Phys. J. E}\ }\textbf {\bibinfo {volume} {46}},\
  \bibinfo {pages} {10} (\bibinfo {year} {2023})}\BibitemShut {NoStop}%
\bibitem [{\citenamefont {Duval}\ \emph {et~al.}(2023)\citenamefont {Duval},
  \citenamefont {Schmidt}, \citenamefont {Hern{\'a}ndez-Garc{\'\i}a},
  \citenamefont {Miret}, \citenamefont {Malliaros}, \citenamefont {Bengio},\
  and\ \citenamefont {Rolnick}}]{duval2023}%
  \BibitemOpen
  \bibfield  {author} {\bibinfo {author} {\bibfnamefont {A.~A.}\ \bibnamefont
  {Duval}}, \bibinfo {author} {\bibfnamefont {V.}~\bibnamefont {Schmidt}},
  \bibinfo {author} {\bibfnamefont {A.}~\bibnamefont
  {Hern{\'a}ndez-Garc{\'\i}a}}, \bibinfo {author} {\bibfnamefont
  {S.}~\bibnamefont {Miret}}, \bibinfo {author} {\bibfnamefont {F.~D.}\
  \bibnamefont {Malliaros}}, \bibinfo {author} {\bibfnamefont {Y.}~\bibnamefont
  {Bengio}},\ and\ \bibinfo {author} {\bibfnamefont {D.}~\bibnamefont
  {Rolnick}},\ }\bibfield  {title} {\bibinfo {title} {{FAENet}: Frame averaging
  equivariant {GNN} for materials modeling},\ }in\ \href
  {https://proceedings.mlr.press/v202/duval23a.html} {\emph {\bibinfo
  {booktitle} {Proceedings of the 40th International Conference on Machine
  Learning}}},\ \bibinfo {series} {Proceedings of Machine Learning Research},
  Vol.\ \bibinfo {volume} {202}\ (\bibinfo  {publisher} {PMLR},\ \bibinfo
  {year} {2023})\ pp.\ \bibinfo {pages} {9013--9033},\ \Eprint
  {https://arxiv.org/abs/2305.05577} {arXiv:2305.05577 [cs.LG]} \BibitemShut
  {NoStop}%
\bibitem [{\citenamefont {Kohn}(1996)}]{kohn96}%
  \BibitemOpen
  \bibfield  {author} {\bibinfo {author} {\bibfnamefont {W.}~\bibnamefont
  {Kohn}},\ }\bibfield  {title} {\bibinfo {title} {Density functional and
  density matrix method scaling linearly with the number of atoms},\ }\href
  {https://doi.org/10.1103/PhysRevLett.76.3168} {\bibfield  {journal} {\bibinfo
   {journal} {Phys. Rev. Lett.}\ }\textbf {\bibinfo {volume} {76}},\ \bibinfo
  {pages} {3168} (\bibinfo {year} {1996})}\BibitemShut {NoStop}%
\bibitem [{\citenamefont {Prodan}\ and\ \citenamefont {Kohn}(2005)}]{prodan05}%
  \BibitemOpen
  \bibfield  {author} {\bibinfo {author} {\bibfnamefont {E.}~\bibnamefont
  {Prodan}}\ and\ \bibinfo {author} {\bibfnamefont {W.}~\bibnamefont {Kohn}},\
  }\bibfield  {title} {\bibinfo {title} {Nearsightedness of electronic
  matter},\ }\href {https://doi.org/10.1073/pnas.0505436102} {\bibfield
  {journal} {\bibinfo  {journal} {Proceedings of the National Academy of
  Sciences}\ }\textbf {\bibinfo {volume} {102}},\ \bibinfo {pages} {11635}
  (\bibinfo {year} {2005})}\BibitemShut {NoStop}%
\bibitem [{\citenamefont {Cybenko}(1989)}]{cybenko89}%
  \BibitemOpen
  \bibfield  {author} {\bibinfo {author} {\bibfnamefont {G.}~\bibnamefont
  {Cybenko}},\ }\bibfield  {title} {\bibinfo {title} {Approximation by
  superpositions of a sigmoidal function},\ }\href
  {https://doi.org/10.1007/BF02551274} {\bibfield  {journal} {\bibinfo
  {journal} {Mathematics of Control, Signals and Systems}\ }\textbf {\bibinfo
  {volume} {2}},\ \bibinfo {pages} {303} (\bibinfo {year} {1989})}\BibitemShut
  {NoStop}%
\bibitem [{\citenamefont {Hornik}\ \emph {et~al.}(1989)\citenamefont {Hornik},
  \citenamefont {Stinchcombe},\ and\ \citenamefont {White}}]{hornik89}%
  \BibitemOpen
  \bibfield  {author} {\bibinfo {author} {\bibfnamefont {K.}~\bibnamefont
  {Hornik}}, \bibinfo {author} {\bibfnamefont {M.}~\bibnamefont
  {Stinchcombe}},\ and\ \bibinfo {author} {\bibfnamefont {H.}~\bibnamefont
  {White}},\ }\bibfield  {title} {\bibinfo {title} {Multilayer feedforward
  networks are universal approximators},\ }\href
  {https://doi.org/10.1016/0893-6080(89)90020-8} {\bibfield  {journal}
  {\bibinfo  {journal} {Neural Networks}\ }\textbf {\bibinfo {volume} {2}},\
  \bibinfo {pages} {359} (\bibinfo {year} {1989})}\BibitemShut {NoStop}%
\bibitem [{\citenamefont {Barron}(1993)}]{barron93}%
  \BibitemOpen
  \bibfield  {author} {\bibinfo {author} {\bibfnamefont {A.~R.}\ \bibnamefont
  {Barron}},\ }\bibfield  {title} {\bibinfo {title} {Universal approximation
  bounds for superpositions of a sigmoidal function},\ }\href
  {https://doi.org/10.1109/18.256500} {\bibfield  {journal} {\bibinfo
  {journal} {IEEE Transactions on Information Theory}\ }\textbf {\bibinfo
  {volume} {39}},\ \bibinfo {pages} {930} (\bibinfo {year} {1993})}\BibitemShut
  {NoStop}%
\bibitem [{\citenamefont {Falicov}\ and\ \citenamefont
  {Kimball}(1969)}]{falicov69}%
  \BibitemOpen
  \bibfield  {author} {\bibinfo {author} {\bibfnamefont {L.~M.}\ \bibnamefont
  {Falicov}}\ and\ \bibinfo {author} {\bibfnamefont {J.~C.}\ \bibnamefont
  {Kimball}},\ }\bibfield  {title} {\bibinfo {title} {Simple model for
  semiconductor-metal transitions: {SmB$_6$} and transition-metal oxides},\
  }\href {https://doi.org/10.1103/PhysRevLett.22.997} {\bibfield  {journal}
  {\bibinfo  {journal} {Phys. Rev. Lett.}\ }\textbf {\bibinfo {volume} {22}},\
  \bibinfo {pages} {997} (\bibinfo {year} {1969})}\BibitemShut {NoStop}%
\bibitem [{\citenamefont {Kennedy}\ and\ \citenamefont
  {Lieb}(1986)}]{kennedy86}%
  \BibitemOpen
  \bibfield  {author} {\bibinfo {author} {\bibfnamefont {T.}~\bibnamefont
  {Kennedy}}\ and\ \bibinfo {author} {\bibfnamefont {E.~H.}\ \bibnamefont
  {Lieb}},\ }\bibfield  {title} {\bibinfo {title} {An itinerant electron model
  with crystalline or magnetic long range order},\ }\href
  {https://doi.org/10.1016/0378-4371(86)90188-3} {\bibfield  {journal}
  {\bibinfo  {journal} {Physica A}\ }\textbf {\bibinfo {volume} {138}},\
  \bibinfo {pages} {320} (\bibinfo {year} {1986})}\BibitemShut {NoStop}%
\bibitem [{\citenamefont {Freericks}\ and\ \citenamefont
  {Zlati{\'c}}(2003)}]{freericks03}%
  \BibitemOpen
  \bibfield  {author} {\bibinfo {author} {\bibfnamefont {J.~K.}\ \bibnamefont
  {Freericks}}\ and\ \bibinfo {author} {\bibfnamefont {V.}~\bibnamefont
  {Zlati{\'c}}},\ }\bibfield  {title} {\bibinfo {title} {Exact dynamical
  mean-field theory of the {Falicov-Kimball} model},\ }\href
  {https://doi.org/10.1103/RevModPhys.75.1333} {\bibfield  {journal} {\bibinfo
  {journal} {Rev. Mod. Phys.}\ }\textbf {\bibinfo {volume} {75}},\ \bibinfo
  {pages} {1333} (\bibinfo {year} {2003})}\BibitemShut {NoStop}%
\bibitem [{\citenamefont {Freericks}\ and\ \citenamefont
  {Lema{\'n}ski}(2000)}]{freericks00}%
  \BibitemOpen
  \bibfield  {author} {\bibinfo {author} {\bibfnamefont {J.~K.}\ \bibnamefont
  {Freericks}}\ and\ \bibinfo {author} {\bibfnamefont {R.}~\bibnamefont
  {Lema{\'n}ski}},\ }\bibfield  {title} {\bibinfo {title} {Segregation and
  charge-density-wave order in the spinless {Falicov-Kimball} model},\ }\href
  {https://doi.org/10.1103/PhysRevB.61.13438} {\bibfield  {journal} {\bibinfo
  {journal} {Phys. Rev. B}\ }\textbf {\bibinfo {volume} {61}},\ \bibinfo
  {pages} {13438} (\bibinfo {year} {2000})}\BibitemShut {NoStop}%
\bibitem [{\citenamefont {Freericks}\ \emph {et~al.}(2002)\citenamefont
  {Freericks}, \citenamefont {Lieb},\ and\ \citenamefont
  {Ueltschi}}]{freericks02}%
  \BibitemOpen
  \bibfield  {author} {\bibinfo {author} {\bibfnamefont {J.~K.}\ \bibnamefont
  {Freericks}}, \bibinfo {author} {\bibfnamefont {E.~H.}\ \bibnamefont
  {Lieb}},\ and\ \bibinfo {author} {\bibfnamefont {D.}~\bibnamefont
  {Ueltschi}},\ }\bibfield  {title} {\bibinfo {title} {Phase separation due to
  quantum mechanical correlations},\ }\href
  {https://doi.org/10.1103/PhysRevLett.88.106401} {\bibfield  {journal}
  {\bibinfo  {journal} {Phys. Rev. Lett.}\ }\textbf {\bibinfo {volume} {88}},\
  \bibinfo {pages} {106401} (\bibinfo {year} {2002})}\BibitemShut {NoStop}%
\bibitem [{\citenamefont {Lema{\'n}ski}\ \emph {et~al.}(2002)\citenamefont
  {Lema{\'n}ski}, \citenamefont {Freericks},\ and\ \citenamefont
  {Banach}}]{lemanski02}%
  \BibitemOpen
  \bibfield  {author} {\bibinfo {author} {\bibfnamefont {R.}~\bibnamefont
  {Lema{\'n}ski}}, \bibinfo {author} {\bibfnamefont {J.~K.}\ \bibnamefont
  {Freericks}},\ and\ \bibinfo {author} {\bibfnamefont {G.}~\bibnamefont
  {Banach}},\ }\bibfield  {title} {\bibinfo {title} {Stripe phases in the
  two-dimensional {Falicov-Kimball} model},\ }\href
  {https://doi.org/10.1103/PhysRevLett.89.196403} {\bibfield  {journal}
  {\bibinfo  {journal} {Phys. Rev. Lett.}\ }\textbf {\bibinfo {volume} {89}},\
  \bibinfo {pages} {196403} (\bibinfo {year} {2002})}\BibitemShut {NoStop}%
\bibitem [{\citenamefont {Fan}\ and\ \citenamefont {Chern}(2026)}]{Fan2026b}%
  \BibitemOpen
  \bibfield  {author} {\bibinfo {author} {\bibfnamefont {Y.}~\bibnamefont
  {Fan}}\ and\ \bibinfo {author} {\bibfnamefont {G.-W.}\ \bibnamefont
  {Chern}},\ }\bibfield  {title} {\bibinfo {title} {Graph neural network force
  fields for adiabatic dynamics of lattice hamiltonians},\ }\href
  {https://doi.org/10.1063/5.0334142} {\bibfield  {journal} {\bibinfo
  {journal} {APL Machine Learning}\ }\textbf {\bibinfo {volume} {4}},\ \bibinfo
  {pages} {036107} (\bibinfo {year} {2026})}\BibitemShut {NoStop}%
\end{thebibliography}%

\end{document}